\documentclass[preprint]{aastex}
\usepackage{graphicx}
\usepackage{natbib}
\begin{document}

\title{The shape and dynamics of a heliotropic dusty ringlet \\
in the Cassini Division}
\author{M.M. Hedman$^{a}$, J.A. Burt$^{a,b}$, J.A. Burns$^{a,c}$, M.S. Tiscareno$^a$}
\affil{$^a$ Department of Astronomy, Cornell Unviersity, Ithaca NY 14853}
\affil{$^b$ Department of Astronomy and Astrophysics, University of California, Santa Cruz, CA 95064}
\affil{$^c$ College of Engineering, Cornell University, Ithaca NY 14853}

\bigskip


\bigskip

\noindent \\
Corresponding Author: Matthew Hedman\\
Space Sciences Building\\
Cornell University\\
Ithaca NY 14853\\
607-255-5913\\
mmhedman@astro.cornell.edu

\pagebreak

The so-called ``Charming Ringlet'' (R/2006 S3) is a low-optical-depth, dusty ringlet
located in the Laplace gap in the Cassini Division, roughly 119,940 km 
from Saturn center.  This ringlet is particularly interesting because
its radial position varies systematically with longitude relative to the Sun
in such a way that the ringlet's geometric center appears to be displaced
away from Saturn's center in a direction roughly toward the Sun.
In other words, the ringlet is always found at greater distances
from the planet's center at longitudes near the sub-solar longitude
than it is at longitudes near Saturn's shadow.
This ``heliotropic'' behavior indicates that the dynamics of the particles in this ring
are being influenced by solar radiation pressure.  In order to investigate this 
phenomenon,  which has been predicted theoretically but not
observed this clearly, we analyze multiple image sequences of this ringlet obtained
by the Cassini spacecraft in order to constrain its shape and orientation.
These data can be fit reasonably well with a 
model in which both the eccentricity and the inclination of the
ringlet have  ``forced" components (that maintain a 
fixed orientation relative to the Sun) as well as ``free" components
(that drift around the planet at steady rates determined
by Saturn's oblateness). The best-fit value for the
eccentricity forced by the Sun is $0.000142\pm0.000004$, assuming
this component of the eccentricity has its pericenter
perfectly anti-aligned with the Sun. These data also place an
upper limit on a forced inclination of 0.0007$^\circ$. Assuming the forced
inclination is zero and the forced eccentricity vector is aligned 
with the anti-solar direction, the best-fit values for the free 
components of the eccentricity and inclination are $0.000066\pm0.000003$ 
and $0.0014\pm0.0001^\circ$,
respectively. While the magnitude of the forced eccentricity
is roughly consistent with theoretical expectations for radiation
pressure acting on 10-to-100-micron-wide icy grains, the 
existence of significant free eccentricities and inclinations
poses a significant challenge for models of low-optical-depth dusty rings.

\section{Introduction}

\begin{figure}
\centerline{\resizebox{5.5in}{!}{\includegraphics{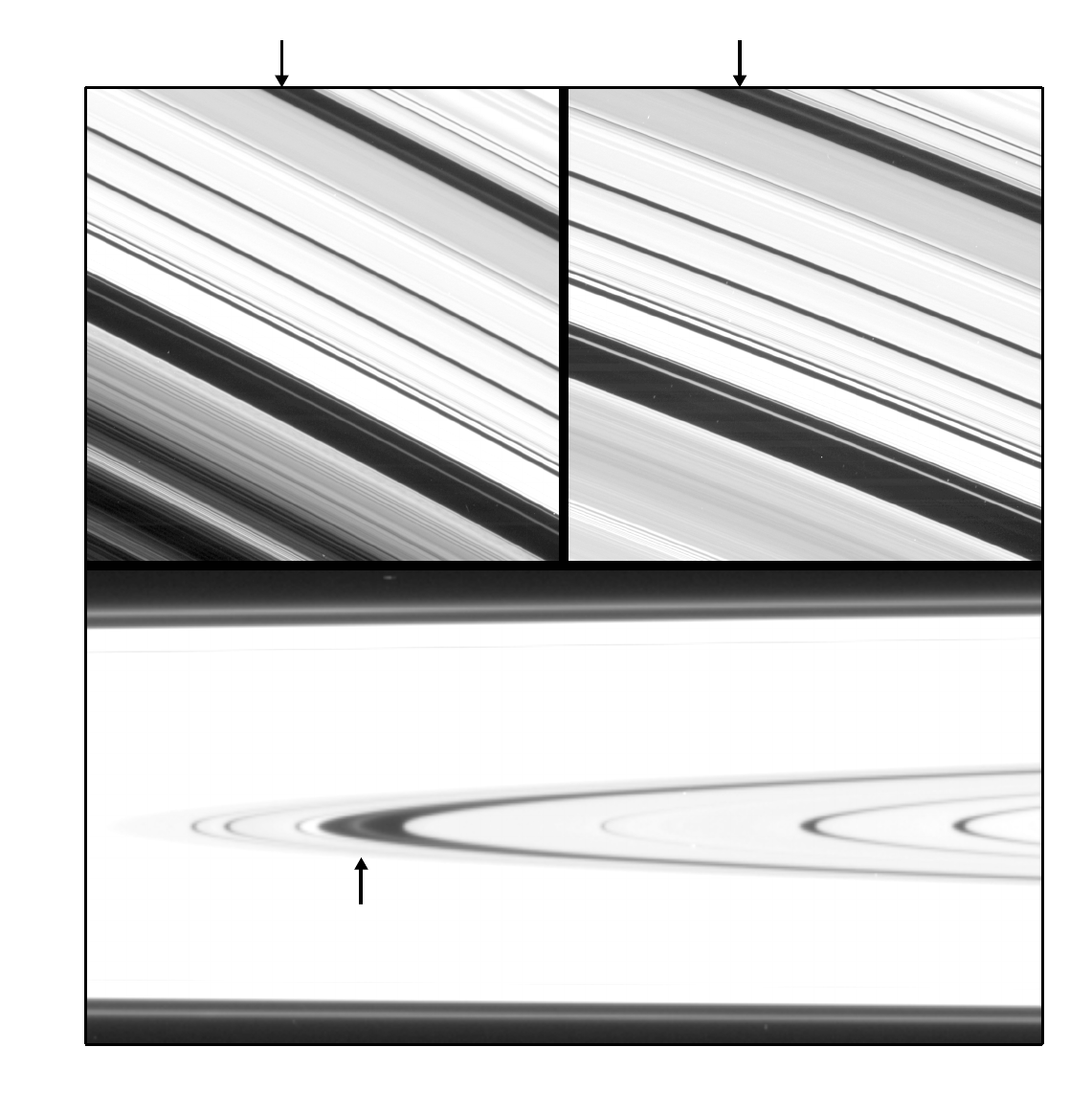}}}
\caption{Sample images of the Charming Ringlet in the Cassini Division 
obtained by the narrow-angle  camera onboard the Cassini spacecraft.
The top two images were obtained on day 343 of 2008 as part
of the RETARMRMP observation in Orbit 96, when the sub-solar
longitude was 217$^\circ$ (see Table~\ref{obstab1}).
The two images have been separately cropped, rotated and stretched
to facilitate comparisons. In both images, radius in the rings increases
towards to upper right. The arrows at the top of the image point 
to the Charming Ringlet in the Laplace gap. Note that in the 
left-hand image (N1607440846,  observed longitude=5$^\circ$) of a region
near Saturn's shadow, the ringlet is closer to the inner edge of the 
gap, while in the right-hand  image (N1609443806, 
observed longitude=$192^\circ$) of a region near to the 
sub-solar longitude, the ringlet is closer to the outer edge of the gap.
The bottom image (N1547759879) was obtained on day 17 of 2007
as part of the RPXMOVIE observation in Orbit 37, when the ring opening
angle was only -0.36$^\circ$. The image has been rotated so that Saturn's
north pole points upwards. Ring radius increases from right to left, and
the arrow points to the Charming Ringlet in the Laplace Gap. Note
that the ringlet appears slightly displaced upwards in this image
relative to the edges of the gap (the upper arm of the ring disappears into the glare of the edge of
the gap faster than the lower arm). This suggests that this ringlet is
inclined.}
\label{charm_im}
\end{figure}

Images taken by the cameras onboard the
Cassini spacecraft have revealed that several
of the wider gaps in Saturn's main rings contain
low-optical-depth, dusty ringlets \citep{Porco05}. One of these
ringlets is located in the 200-km wide space 
in the outer Cassini Division between
the inner edges of the Laplace Gap and the
Laplace Ringlet, 119,940 km from Saturn's center. 
This ringlet has a peak normal optical depth 
of around $10^{-3}$ and its photometric properties
(such as a dramatic increase in brightness at
high phase angles) indicate that it is composed primarily
of small dust grains less than $100$ microns across \citep{Horanyi09}. 
While this feature is officially designated R/2006 S3 \citep{Porco06}, 
it is unofficially called the ``Charming Ringlet"
by various Cassini scientists, and we will use that name here.  
Regardless of its name, this ringlet is of special
interest because its radial position varies 
systematically with longitude relative to the Sun
in such a way that the ringlet's geometric center 
appears to be displaced away from Saturn's center 
towards the Sun. In other words, this ringlet
always appears some tens of
kilometers further from the planet's center at longitudes near 
the sub-solar longitude than it is at longitudes near Saturn's shadow
(see Figure~\ref{charm_im}).
This ``heliotropic'' behavior suggests that
non-gravitational forces such as solar radiation
pressure are affecting the particles' orbital dynamics, as predicted
by various theoretical models (e.g. Hor{\'a}nyi and Burns 1991, Hamilton 1993). 

While other dusty ringlets, like those in the Encke Gap,
may also show heliotropic behavior \citep{Hedman07},
the Charming Ringlet provides the best opportunity
to begin investigations of this phenomenon. Unlike the Encke Gap
ringlets, the Charming Ringlet does not appear to contain
bright clumps or noticeable short-wavelength ``kinks" in its radial position. The
absence of such features makes the global shape of the
ringlet easier to observe and quantify. Furthermore,
the radial positions of the edges of the Laplace gap and ringlet only
vary by a few kilometers \citep{Hedman10}, so this gap is a much simpler environment
than other gaps (like the Huygens gap) where the radial locations of the edges
can vary by tens of kilometers. Finally, the observations of
the Charming Ringlet are more extensive than those of some other
dusty ringlets. 

In this paper, we build upon the preliminary work 
reported in ~\citet{Hedman07} and ~\citet{Burt08}
in order to develop a model for the three-dimensional 
shape and orientation of the Charming Ringlet
and to explore what such a model implies about
the particle dynamics in this ring. First, we provide
a brief summary of the data that will be used in this
analysis and then fit the different data sets
to models of an eccentric, inclined ringlet. These
fits indicate that the shape and orientation of the
ringlet change significantly over time.
Next, we review the
theoretical predictions for how particle orbits should behave
under the influence of solar radiation pressure. 
Based on this theory, we 
develop a global model that includes both forced
and free components in the ringlet's 
eccentricity and inclination; these can reproduce
the observations reasonably well. Finally, we 
discuss the implications of such a model for the dynamics
of this ringlet.

\section{Observations and data reduction}

\begin{table}
\caption{Data sets used in this analysis}
\label{obstab1}

(a) Longitudinal-scan observations used in this analysis

\resizebox{6in}{!}{\begin{tabular}{|l|c|c|c|c|c|c|c|c|c|}\hline
Orbit/Obs. Sequence & Date & Images & Radial & Phase & Elevation & Sub-Spacecraft & Subsolar & Subsolar & Observed \\
& & & resolution &  &  & longitude & latitude &  longitude & longitude \\ \hline
030/AZDKMRHPH001/ & 2006-290 & N1539746533-N1539760916 (59) & 12-21 km/pixel &
153-157$^\circ$ & 35.5-40.1$^\circ$ & 357.4-358.4$^\circ$ 
&  -15.4$^\circ$  & 191.3$^\circ$ & 104.8-341.9$^\circ$\\
042/RETMDRESA001/ & 2007-099 & N1554850367-N1554852347 (10) & 5-9 km/pixel &
7-19$^\circ$ & (-21.3)-(-18.7)$^\circ$ & 205.8-206.4$^\circ$ 
& -13.0$^\circ$ & 197.3$^\circ$ & 67.2-342.4$^\circ$\\
070/RETMDRESA001/ & 2008-151 & N1590852409-N1590863423 (46) & 7-12 km/pixel &
30-35$^\circ$ & 21.4-26.9$^\circ$ & 201.6-202.1$^\circ$ 
& -6.7$^\circ$ & 211.0$^\circ$ & 132.4-350.1$^\circ$ \\
071/PAZSCN002/          & 2008-159 & N1591528309-N1591548903 (93) & 5-9 km/pixel &
41-44$^\circ$ & 34.2-37.5$^\circ$ & 204.7-206.3$^\circ$ 
& -6.6$^\circ$ & 211.2$^\circ$ & 91.5-233.4$^\circ$ \\
082/RETARMRLP001/ & 2008-237 & N1598292162-N1598301617 (40) & 6-10 km/pixel & 
36-45$^\circ$ & 30.9-38.8$^\circ$ & 204.2-205.0$^\circ$ 
& -5.4$^\circ$ & 213.7$^\circ$ & 67.0-359.0$^\circ$ \\
092/RETARMRLF001/ & 2008-312 & N1604731407-N1604737767 (39) & 5-8 km/pixel &
46-56$^\circ$ & 41.9-52.1$^\circ$ & 214.3-215.4$^\circ$  
& -4.3$^\circ$ & 216.0$^\circ$ &72.1-357.1$^\circ$ \\
096/RETARMRMP001/ & 2008-343 & N1607440846-N1607445736 (26) & 4-6 km/pixel &
57-74$^\circ$ & 53.9-70.1$^\circ$ & 224.9-228.1$^\circ$ 
& -3.8$^\circ$ & 217.0$^\circ$ & 77.4-364.7$^\circ$ \\
\hline
\end{tabular}}

{(b)Elevation-scan observations used in this analysis}

\resizebox{6in}{!}{\begin{tabular}{|l|c|c|c|c|c|c|c|c|c|}\hline
Orbit/Obs. Sequence &  Date& Images & Radial & Phase & Elevation & Sub-Spacecraft & Subsolar & Subsolar & Observed \\
& & & resolution & &  & longitude & latitude &  longitude & longitude \\ \hline
037/RPXMOVIE001/  & 2007-017 & N1547739487-N1547782929 (34) & 5-6 km/pixel &
32-47$^\circ$ & (-8.7)-(+8.1)$^\circ$ & 218.6-228.5$^\circ$ & 
-14.1$^\circ$ & 194.5$^\circ$ &  130-150$^\circ$ \\
\hline
\end{tabular}}
\end{table}

All the images used in this analysis were obtained by the 
Narrow-Angle Camera (NAC) of the Imaging Science
Subsystem (ISS) onboard the Cassini spacecraft \citep{porco04}.
While ISS has obtained many images of the Charming Ringlet
over the course of the Cassini mission,
we will focus here exclusively on a limited sub-set of these 
data from a few imaging sequences. 
Each of these sequences was obtained over a relatively
short period of time and covers a sufficient range of longitudes 
or viewing geometries that it can provide useful constraints 
on the shape and  orientation of the ring.  These data sets
are therefore particularly useful for developing a 
shape model for this ring. In principle, once a rough model
has been established, additional data can be used to
refine the model parameters and test the model. However,
such an analysis is beyond the scope of this paper
and therefore will be the subject of future work.

Two different types of observation sequences will
be utilized in the present study,  
``longitudinal scans" and ``elevation scans". 
Each longitudinal scan consists of a series of images of the Cassini
Division, with different images centered at different inertial longitudes
in the rings. These scans provide
maps of the apparent radial position of the Charming Ringlet 
as a function of longitude relative to the Sun. The
seven such scans used in this analysis (listed in Table~\ref{obstab1}a)
are all the scans obtained prior to 2009 that contain the Charming Ringlet, have
sufficient radial resolution to clearly resolve the ringlet and
also  cover a sufficiently broad range of longitudes
($>140^\circ$) to provide a reliable measurement of both
the ringlet's eccentricity and inclination (see below). 

By contrast, elevation scans consist of a series of images
of the ring ansa taken over a period of time 
when the spacecraft passed through the
ring-plane, yielding observations covering a range of 
ring-opening angles  $B$ around zero. 
Such images provide limited information about the
ringlet's eccentricity, however, observable shifts in the
ringlet's apparent position relative to other 
ring features  provide evidence that the ringlet is inclined 
(see Fig~\ref{charm_im}). These observations therefore can
furnish additional constraints on the ringlet's vertical structure.
Thus far, only one image sequence (given in Table~\ref{obstab1}b)
has sufficient resolution and elevation-angle coverage to
yield useful constraints on the ringlet's orientation.

All of these images were processed using the standard 
CISSCAL calibration routines (version 3.6) \citep{porco04} that remove backgrounds,
flat-field the images, and convert the raw data numbers 
into $I/F$ (a standardized measure of reflectance where $I$ is
the intensity of the scattered radiation while $\pi F$ is the
solar flux at Saturn). \nocite{French93}
We then extracted measurements of the
ringlet's radial position with the following procedures.

First, all the relevant images were geometrically navigated employing
the appropriate SPICE kernels to establish the position and 
approximate pointing of the spacecraft. The pointing was refined
using the outer edge of the Jeffreys Gap (called OEG 15 in  French {\it et al.} 1993, 
assumed to be circular and lie at 118,968 km) as a fiducial feature. 
Recent Cassini occultation measurements demonstrate that
this feature is circular to better than 1 km \citep{Hedman10,French10},
making it a reliable reference point in the rings.

Once each image was navigated,
the brightness data were converted into radial brightness profiles
by averaging the brightness at each radius over a range of longitudes.
For the longitudinal scans, each image covered a sufficiently small range of 
longitudes that variations in the radial position of the ringlet
within an image could be ignored. Consequently, a single radial
scan was derived from each image by averaging the data over
all observed longitudes. By contrast, for the elevation scans,
variations in the radial position of the ringlet were apparent within
individual images. A series of 8-20 radial brightness profiles was therefore
extracted from each image, with each profile being the average
brightness of the ring in a range of longitudes  between 
0.5$^\circ$ and 1.0$^\circ$ wide. Note that for all these profiles, 
the radius scale corresponds to the projected position of any given
feature onto the ring-plane.

\begin{figure}
\resizebox{6in}{!}{\includegraphics{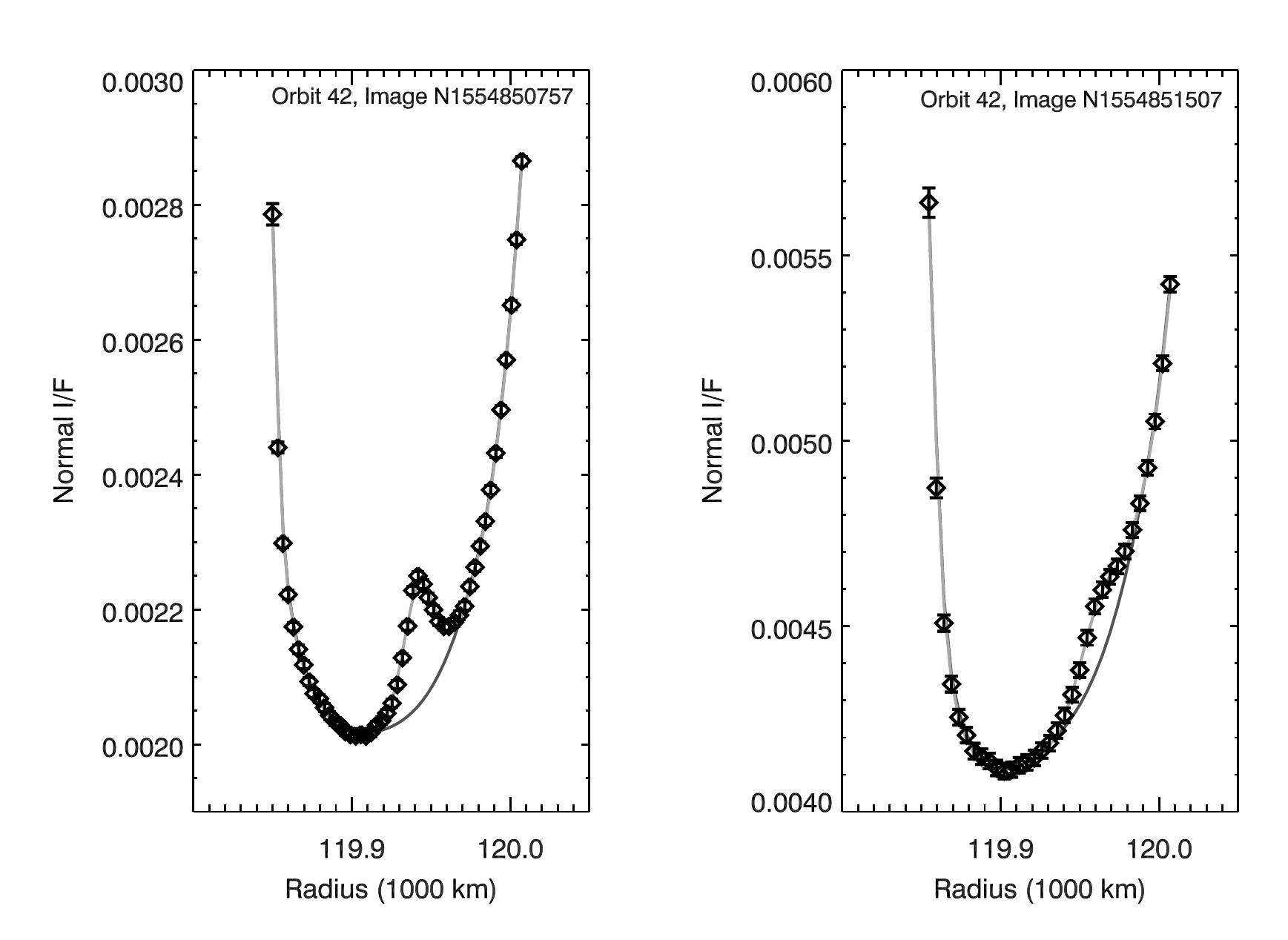}}
\caption{Examples of the profile fitting procedures described in the text.
In each plot, the data points shows the brightness profile across the Laplace Gap, including
the Charming Ringlet. The data are given in terms of  `Normal $I/F$', which is
the observed $I/F$ multiplied by the sine of the ring opening angle.
The dark grey curve shows the fit background profile,
while the light grey curve  shows this background plus the best-fit gaussian
profile for the ringlet. Note the example on the right is among the most
extreme in terms of the subtleness of the ring signal, and even here the fit
is very good. Most of the fits used in this analysis are more like the example on the left.}
\label{prof}
\end{figure}

The Charming Ringlet could be detected as a brightness peak within
the Laplace gap in all of these radial scans. The radial position of
the ringlet was estimated from each scan by 
fitting the ringlet's brightness profile to a Gaussian. For the high-phase
observations in Orbit 30, the ringlet was sufficiently bright that
the Gaussian could be fit directly to the radial profile. For the 
other (lower-phase) profiles, however, the ringlet was considerably
fainter and the brightness variations within the gap due to various
instrumental effects could not be ignored. In these situations, a background light
profile for the gap was computed using the data outside the ringlet
(the edges of the ringlet were determined based on where the slope
of the brightness profile around the ringlet was closest to zero). This background was interpolated
into the region under the ringlet (using a  spline interpolation of the profile smoothed over
three radial bins) and a Gaussian was fit to the background-subtracted
ringlet profile. Figure~\ref{prof} shows examples of the raw profile, the interpolated
background and the background plus the fitted Gaussian, demonstrating that
this procedure yields sensible results even when the ringlet is rather subtle.

\begin{figure}
\centerline{\resizebox{5.5in}{!}{\includegraphics{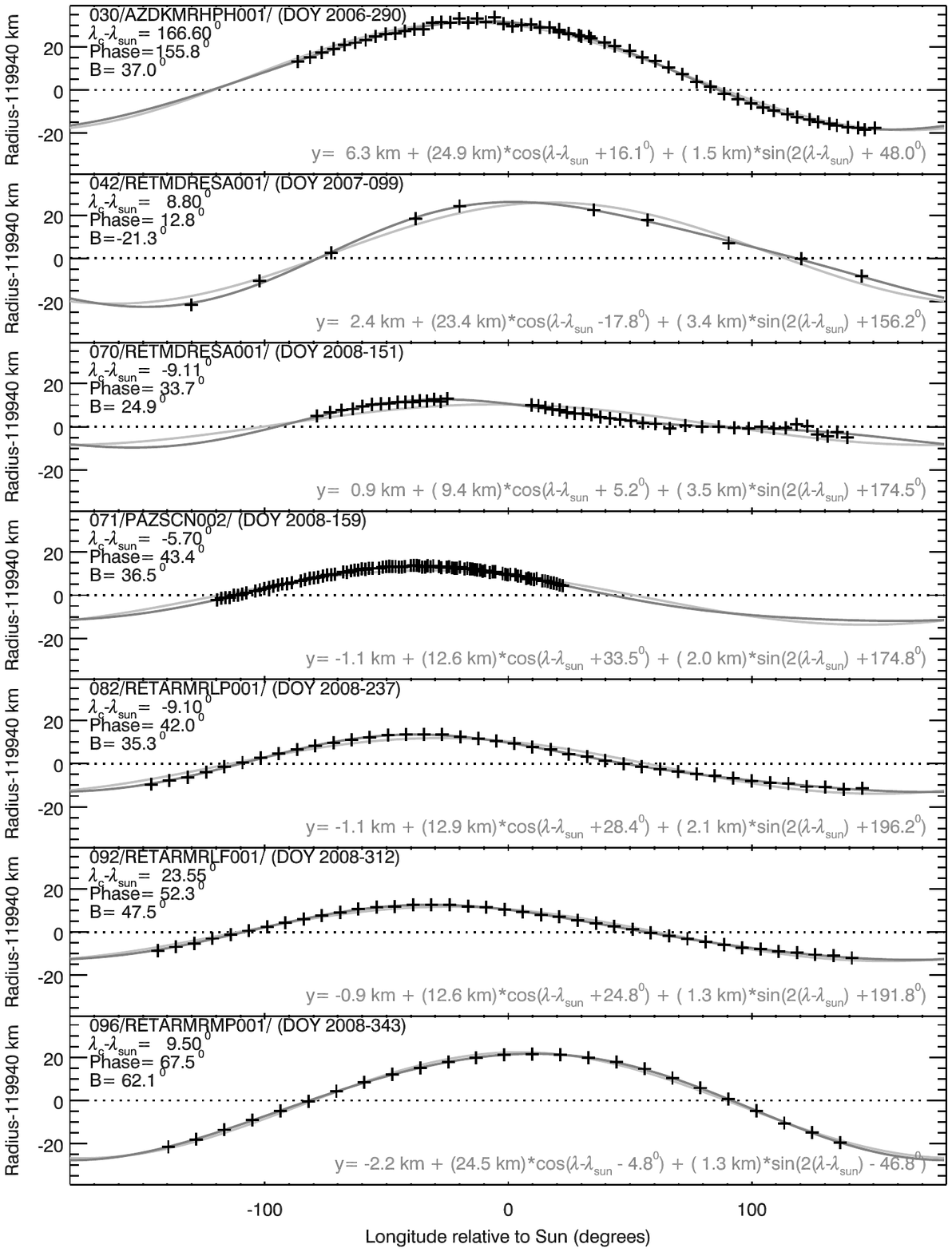}}}
\caption{The apparent radius of the Charming Ringlet (projected
onto the ring-plane) as a function of longitude relative to the Sun,
derived from the seven longitudinal scans. The observations
are shown as crosses. The dark grey curve
shows the best-fit model to each data set with the parameters
listed on each plot  (compare with Eq.~\ref{shape}). 
The light grey curve is the same model with
the term $\propto \sin(2\lambda)$ removed to illustrate
the importance of this term to the overall fit.}
\label{longscans}
\end{figure}

The above process yielded a series of measurements
of the apparent radial position of the ringlet as a function
of longitude.  Figure~\ref{longscans} shows these data for the
seven different longitudinal scans. Note that in all cases
the ringlet is found furthest from the planet at a point near to the
sub-solar longitude. This is not just a coincidence of when the
ringlet was observed, but is instead the evidence
for the ``heliotropic'' character of this ringlet. However, we
can also observe that the apparent shape of the ringlet 
varies significantly among the different observations.
This implies that the ringlet does not simply maintain
a fixed orientation relative to the Sun, but instead
has a more complex and time-variable shape.

\begin{figure}
\resizebox{6in}{!}{\includegraphics{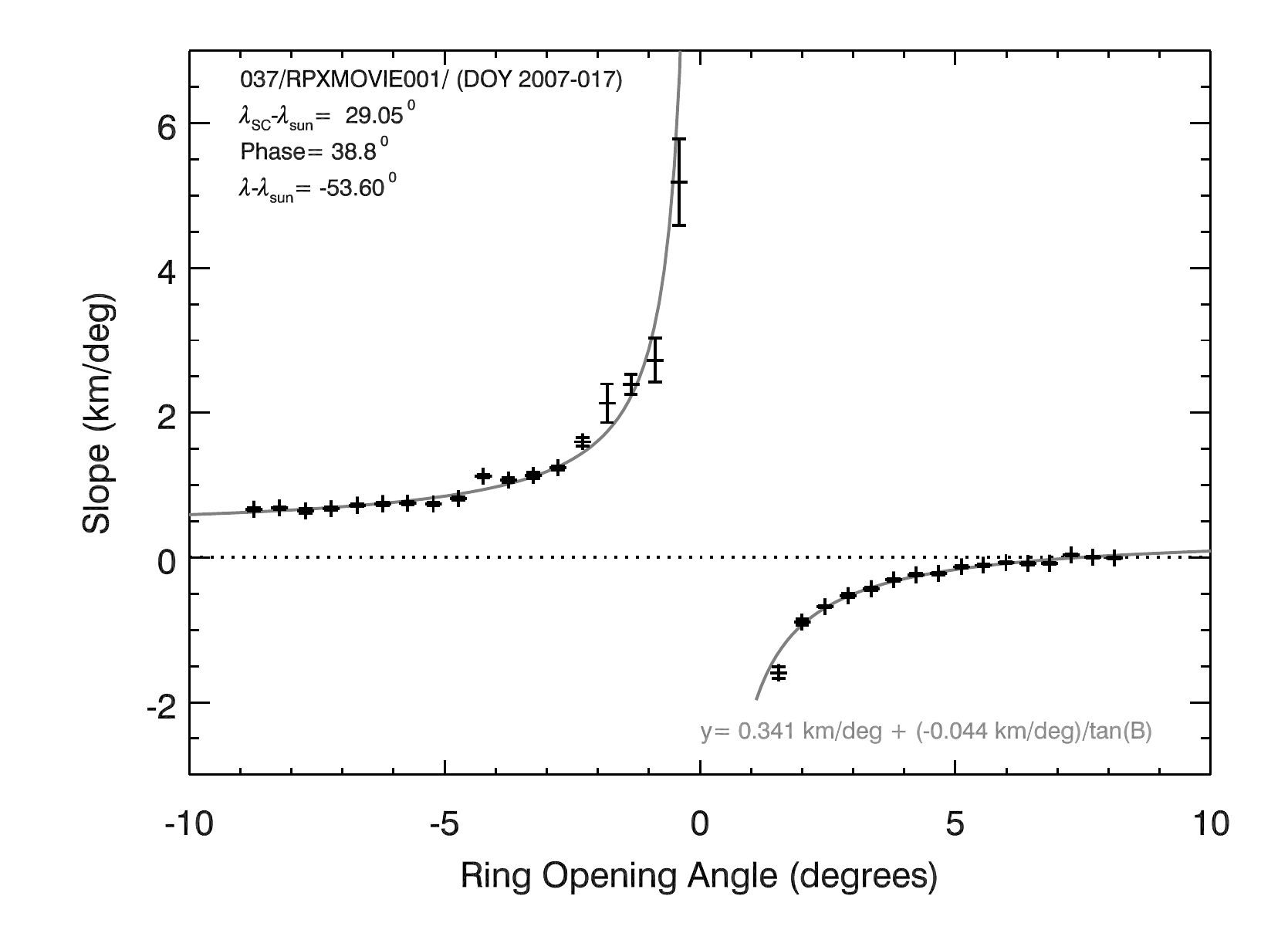}}
\caption{The slope in the apparent radius versus longitude, plotted
against ring-opening angle $B$ to the spacecraft, as derived from the
single elevation scan.  The curve is the best-fitting model with the
parameters shown (compare with Eq.~\ref{slopeshape}).}
\label{elscan}
\end{figure}

For the elevation scan, the radial position of the ringlet versus
longitude from each image can be fit to a line. Figure~\ref{elscan} shows
the slopes of the line derived from these images as a function
of ring-opening angle $B$. The slope changes dramatically
as the spacecraft crosses the ringplane. This strongly suggests
that this portion of the ring is vertically displaced from the ringplane
\citep{Burt08}, and means that we will need a three-dimensional
model to fully describe the shape of this ringlet.

\section{Ringlet shape estimates from individual observations}

The above evidence for time-variable and three-dimensional
structure obviously complicates our efforts to quantify 
the Charming Ringlet's shape. Fortunately, it turns
out that the data from individual scans can be reasonably
well fit by simple models of eccentric, inclined ringlets.
By fitting each scan to such a model, we can further 
reduce the data to a small number of shape/orbital parameters,
which may change with time.

Each scan consists of measurements of the
apparent radial position of the ringlet projected on the
ringplane $\tilde{r}$ versus longitude relative to the Sun 
$\lambda-\lambda_\Sun=\lambda'$. Assuming the
ringlet can have both an inclination and an eccentricity, the radial and
vertical positions of the ringlet versus longitude are for 
small eccentricities and inclinations well approximated by:
\begin{equation}
r=a-ae\cos(\lambda'-\varpi')
\end{equation}
\begin{equation}
z=ai\sin(\lambda'-\Omega'),
\end{equation}
where $a$, $e$ and $i$ are the semi-major axis, eccentricity and inclination
of the ringlet, and $\varpi'$ and $\Omega'$ are the longitudes of
pericenter and ascending node relative to the Sun.

If $z$ is nonzero, the apparent position of the ringlet will be displaced 
when it is projected onto the ringplane. Assuming the observer is sufficiently
far from the ring, this displacement is simply
\begin{equation}
\delta\tilde{r}=-\frac{z}{\tan B}\cos(\lambda'-\lambda_c'),
\label{rtoz}
\end{equation}
where $B$ is the ring opening angle to the observing spacecraft and
$\lambda'_c$ is the longitude of the spacecraft relative to the Sun.
Substituting in the above value for $z$, we find:
\begin{equation}
\delta\tilde{r}=\frac{-ai}{2\tan B}[\sin(2\lambda'-\Omega'-\lambda_c')-\sin(\Omega'-\lambda'_c)].
\end{equation}
The apparent radial position of such a ringlet is therefore:
\begin{equation}
\tilde{r}=r+\delta\tilde{r}=
a+\frac{ai}{2\tan B}\sin(\Omega'-\lambda'_c)-ae\cos(\lambda'-\varpi')
-\frac{ai}{2\tan B}\sin(2\lambda'-\Omega'-\lambda_c').
\label{shape}
\end{equation}
Note that this expression contains two terms that depend on the longitude $\lambda'$:
one proportional to $e$ and one proportional to $i$. Since these two terms depend
on longitude in different ways, it should be possible to determine both the eccentricity 
and inclination from any observation sequence that covers a sufficiently broad range
of longitudes. Also, since the terms involving $i$ depend on the ring opening angle $B$
while those involving $e$ do not, the effects of inclination and eccentricity 
on the apparent position of the ringlet should also be separable when the 
observation sequences cover a sufficient range in $B$.

\subsection{Elevation Scan}

Over the limited range of longitudes observed in each image of the
elevation scan, the apparent-radius-versus-longitude
curve is well fit by a straight line. Figure~\ref{elscan} shows the slope of this
line as a function of ring opening angle, with error bars derived
from the linear fit. 

Given the above expression (Eq.~\ref{shape}) for the apparent radial position of the 
ring versus longitude, these measured slopes can be identified
with the quantity: 
\begin{equation}
m=\frac{d\tilde{r}}{d\lambda'}=ae\sin(\lambda'-\varpi')
-\frac{ai}{\tan B}\cos(2\lambda'-\Omega'-\lambda_c')
\label{slopeshape}
\end{equation}
In other words, $m=C-z/\tan B$, where $C$ is the constant background slope due
to the eccentricity of the ringlet and $z$ is its vertical displacement
at the observed longitude. Fitting the data from the elevation scan
to an equation of this form, we find that at the observed longitude and time: 
\begin{equation}
z=ai\cos(2\lambda'-\Omega'-\lambda_c') = 2.54 \pm 0.02  \hspace{2pt} {\rm  km},
\label{zeq}
\end{equation}
\begin{equation}
C=ae\sin(\lambda'-\varpi')=19.5 \pm 0.2 \hspace{2pt} {\rm km}.
\label{ceq}
\end{equation}
The curve plotted on Fig.~\ref{elscan} shows this best-fit function, which reproduces
the trends in the data rather well. However, the $\chi^2$ of this fit is 206 for 32
degrees of freedom, indicating that the errors on the individual slope measurements
have been underestimated. Thus the above uncertainties on
$z$ and $C$ should probably be increased by
a factor of 2.5.  Note that while these data alone cannot provide
exact estimates on eccentricity and inclination, we can 
establish that $ai$ is at least 2.5 km and $ae$ is at least 19 km.

\subsection{Longitudinal Scans}

Figure~\ref{longscans} shows the estimated position of the Charming Ringlet versus
longitude relative to the Sun for each of the seven longitudinal scans.
Each of these data sets has been fit to a function of the form (cf Eq.~\ref{shape})
\begin{equation}
\tilde{r}=r_o+r_1\cos(\lambda'-\phi_1)+r_2\cos(2\lambda'-\phi_2).
\end{equation}
The best fit solutions, shown as the dark grey curves in Fig.~\ref{longscans}, satisfactorily
reproduce the trends seen in the real data. We can therefore
use the parameters of this fit and Equation~\ref{shape}
to derive the ring-shape parameters $a,e,i,\varpi'$ and $\Omega'$
(see Table~\ref{longpartab}). The error bars on the orbital parameters 
are computed using the $rms$ residuals from the fit to estimate the 
error bars on each data point. These residuals are always less than one 
kilometer, or about a factor of 10 better than the image resolution 
(see Tables~\ref{obstab1} and~\ref{longpartab}), and probably reflect
small errors and uncertainties in the fitted locations of the fiducial edge
and ringlet center. The small scatter in these data therefore 
confirms the stability of the pointing and fitting algorithms within
each of these sequences.

\begin{table}
\caption{Ring shape parameters derived from the longitudinal scans}
\label{longpartab}
\resizebox{6in}{!}{\begin{tabular}{|l|c|c|c|c|c|c|}\hline
Orbit/Obs. Sequence &  $rms^a$ & $\delta a^b$ & $ae$ & $\varpi-\lambda_\Sun$ 
		& $ai$ & $\Omega-\lambda_\Sun$ \\
 & (km) &  (km) & (km) & (deg) & (km) &(deg) \\ \hline
030/AZDKMRHP001/ & 0.7 & 5.8 & 
24.92$\pm$0.26 & 163.9$\pm$0.7 & 2.27$\pm$0.34 & 325.4$\pm$6.4 \\
042/RETMDRESA001/ & 0.9 & 2.8 &
23.42$\pm$0.52 & 197.8$\pm$0.9 & 2.63$\pm$0.29 & 195.0$\pm$9.7 \\
070/RETMDRESA001/ & 0.9 & -0.4 &
9.42$\pm$0.61 & 174.8$\pm$2.9 & 3.21$\pm$0.35 & 14.6$\pm$5.3 \\
071/PAZSCN002/          & 0.2 & -1.7 &
12.59$\pm$0.95 & 146.5$\pm$1.3 & 3.01$\pm$0.51 & 10.9$\pm$3.6 \\
082/RETARMRLP001/ & 0.3 & -1.1 &
12.88$\pm$0.08 & 151.6$\pm$0.3 & 3.03$\pm$0.09 & 352.9$\pm$2.0 \\
092/RETARMRLF001/ & 0.2 & -0.6 &
12.59$\pm$0.06 & 155.2$\pm$0.2 & 2.75$\pm$0.09 & 349.3$\pm$2.5 \\
 096/RETARMRMP001/  & 0.2  & -1.6 &
 24.55$\pm$0.08 & 184.8$\pm$0.1 & 5.08$\pm$0.24 & 217.3$\pm$2.8 \\
\hline
\end{tabular}}

$^a$ $rms$ residuals of the data after the fit.

$^b$  $a-119940$ km
\end{table}

Let us consider each of these different parameters in turn, starting with the 
semi-major axis $a$. No formal error bars on this parameter
are given here because this parameter is the one most likely 
to be affected by systematic pointing uncertainties between the different scans
caused by differences in the appearance and contrast of
the fiducial edge. Nevertheless,
the scatter in these values is still only a few kilometers and
well below the resolutions of the images (compare to Table~\ref{obstab1}), providing further
confirmation that the fitting procedures employed here are robust.  Note that 
all the  observations give $a$ values within a few kilometers of 119,940 km, which
is very close to exactly halfway between the inner edge of the Laplace Gap
at 119,845 km and the inner edge of the Laplace ringlet at 120,036 km \citep{Hedman10}.

Turning to the eccentricity and pericenter, we may note that while  the pericenter
is always around $180^\circ$ from the Sun, 
both the pericenter location and the eccentricity vary significantly
from one observation to another. The values of $ae$ range from 9.5 km to
25 km, and the pericenter locations deviate from the anti-Sun direction by up to $35^\circ$.
The ringlet therefore does not maintain a perfectly fixed orientation
relative to the Sun. 

Finally, consider the inclination and the node estimates. Six of the seven 
estimates for $ai$  fall in a relatively narrow range of 2.3-3.2 km. The one outlier is the
5.1 km estimate from the Orbit 96 data. However, this observation was made while
the spacecraft was well above the ringplane, and the ring opening angle
changed more over the course of this observation than in any of the others (see Table~\ref{obstab1}).
Thus this measurement of the inclination may be regarded as suspect. Looking at the 
remaining  data, the relatively small scatter in $ai$ may imply that the 
inclination of this ringlet does not vary much with time.
However, we also find that the node positions are very widely scattered. This implies
that this ringlet's line of nodes does not have a fixed orientation relative to the Sun.

\section{Solar radiation pressure and models of heliotropic orbits}

The above data show that the ringlet's pericenter is on average
anti-aligned with the Sun, suggesting  that
a force like solar radiation pressure is influencing the shape
and orientation of this ringlet. However, the eccentricity and alignment
of this ringlet also vary significantly over time, and this indicates that 
the ringlet's dynamics are more complex than we might have expected. 
In order to facilitate the interpretation of these data, 
we will review how solar radiation pressure affects orbital parameters.
This analysis roughly follows
the treatment given in \citet{HB91} for a particle in orbit around Jupiter,
but is generalized to account for the possibility that the Sun may be
located significantly above or below the ringplane. Also, we will restrict
ourselves to nearly circular orbits, thereby  obtaining simpler expressions
than those given by \citet{Hamilton93}. Note that throughout this
analysis we assume the dynamics of the particles is determined
entirely by solar radiation pressure and Saturn's gravity (other
non-gravitational forces such as plasma drag are neglected).

\nocite{Burns76}
We begin with the standard perturbation equations for the semi-major axis $a$, 
eccentricity $e$, inclination $i$, the longitude of periapse $\varpi$
and the longitude of node $\Omega$ of a particle orbit 
 (see e.g. Burns 1976).  Since we are 
interested in orbits with small eccentricities and inclinations, these 
expressions can be approximated as
\begin{equation}
\frac{da}{dt} = {2an}\left[\frac{F_r}{F_{G}}e\sin f
	+\frac{F_t}{F_{G}}(1+e\cos f) \right],
\end{equation}
\begin{equation}
\frac{de}{dt} = n\left[\frac{F_r}{F_{G}}\sin f
	+2\frac{F_t}{F_{G}}\cos f \right],
\label{dedt}
\end{equation}
\begin{equation}
\frac{d\varpi}{dt} = \frac{n}{e}\left[-\frac{F_r}{F_{G}}\cos f 
	+2\frac{F_t}{F_{G}}\sin f\right],
\end{equation}
\begin{equation}
\frac{di}{dt} = n\left[\frac{F_z}{F_{G}}\cos (\varpi-\Omega+f)\right],
\end{equation}
\begin{equation}
\frac{d\Omega}{dt} = \frac{n}{\sin i}\left[\frac{F_z}{F_{G}}\sin (\varpi-\Omega+f)\right],
\label{dOdt}
\end{equation}
where $n$ is the particle's mean motion,
$F_{G}=GMm_g/a^2$ is approximately the force of Saturn's gravity on a particle
with mass $m_g$ (neglecting the effects of Saturn's finite oblateness), $f$ is the particle's true anomaly 
and $F_r$, $F_t$ and  $F_z$ are the radial, azimuthal and normal (to
the orbit plane in the direction of orbital angular momentum) components of the 
perturbing force, respectively.

Say the Sun is located at an elevation angle $B_\Sun$ above the rings
and a longitude $\lambda_\Sun$ in some inertial coordinate system.
Then the components of the solar radiation pressure force $F_\Sun$ at a specified longitude
$\lambda$ in the ring are given by:
\begin{equation}
F_z=-F_\Sun\sin B_\Sun,
\end{equation}
\begin{equation}
F_r=-F_\Sun\cos B_\Sun\cos(\lambda-\lambda_\Sun),
\end{equation}
\begin{equation}
F_t=+F_\Sun\cos B_\Sun\sin(\lambda-\lambda_\Sun).
\end{equation}
Substituting these expressions into the equations of motion, and recognizing that
$f=\lambda-\varpi$, we obtain
\begin{equation}
\frac{da}{dt} = 2an\frac{F_\Sun\cos B_\Sun}{F_{G}}
	[e\sin(\varpi-\lambda_\Sun)+\sin(\lambda-\lambda_\Sun)],
\label{a1}
\end{equation}
\begin{equation}
\frac{de}{dt} = n\frac{F_\Sun\cos B_\Sun}{2F_{G}}
	[3\sin(\varpi-\lambda_\Sun)+\sin(2\lambda-\varpi-\lambda_\Sun)],
\end{equation}
\begin{equation}
\frac{d\varpi}{dt} = \frac{n}{e}\frac{F_\Sun\cos B_\Sun}{2F_{G}}
	[3\cos(\varpi-\lambda_\Sun)-\cos(2\lambda-\varpi-\lambda_\Sun)],
\end{equation}
\begin{equation}
\frac{di}{dt} = -n\frac{F_\Sun\sin B_\Sun}{F_{G}}\cos(\lambda-\Omega),
\end{equation}
\begin{equation}
\frac{d\Omega}{dt} = -\frac{n}{\sin i} \frac{F_\Sun\sin B_\Sun}{F_{G}}\sin(\lambda-\Omega).
\label{o1}
\end{equation}
For small perturbations, we expect that $\varpi$, $\Omega$, $e$, $i$ and $\lambda_\Sun$ will change much more 
slowly than $\lambda$ does. Thus, to obtain the long-term secular evolution of the 
orbital elements, we may average these expression over a single orbit. However, in doing this, 
we must take care to account for Saturn's shadow, which blocks the light
from the Sun during a fraction of the  particle's orbit $\epsilon$. The appropriate
orbit-averaged equations of motion are:
\begin{equation}
\left<\frac{da}{dt}\right> = nae\left[2d(\epsilon)\frac{F_\Sun\cos B_\Sun}{F_{G}}\right]
	\sin(\varpi-\lambda_\Sun),
	\label{a1a}
\end{equation}
\begin{equation}
\left<\frac{de}{dt}\right> = n\left[\frac{3}{2}f(\epsilon)\frac{F_\Sun\cos B_\Sun}{F_{G}}\right]
	\sin(\varpi-\lambda_\Sun),
\end{equation}
\begin{equation}
\left<\frac{d\varpi}{dt}\right> = \frac{n}{e}\left[\frac{3}{2}f(\epsilon)\frac{F_\Sun\cos B_\Sun}{F_{G}}\right]
	\cos(\varpi-\lambda_\Sun),
\end{equation}
\begin{equation}
\left<\frac{di}{dt}\right> = -n\left[g(\epsilon)\frac{F_\Sun\sin B_\Sun}{F_{G}}\right]
	\cos(\Omega-\lambda_\Sun),
\end{equation}
\begin{equation}
\left<\frac{d\Omega}{dt}\right> = \frac{n}{\sin i}\left[g(\epsilon) \frac{F_\Sun\sin B_\Sun}{F_{G}}\right]
	\sin(\Omega-\lambda_\Sun).
	\label{o1a}
\end{equation}
where $d(\epsilon)=1-\epsilon$, $f(\epsilon)=1-\epsilon+\sin(2\pi\epsilon)/6\pi$ and $g(\epsilon)=\sin(\pi\epsilon)/\pi$ (see Appendix).

For an oblate planet like Saturn, these equations of motion are incomplete
because they do not take into account the steady precession in the pericenter
and node caused by  the planet's finite oblateness, which augments the
motion of $\varpi$ and $\Omega$. The full equations of motion 
are therefore: 
\begin{equation}
\left<\frac{da}{dt}\right> = nae\left[2d(\epsilon)\frac{F_\Sun\cos B_\Sun}{F_{G}}\right]
	\sin(\varpi-\lambda_\Sun),
\end{equation}
\begin{equation}
\left<\frac{de}{dt}\right> = n\left[\frac{3}{2}f(\epsilon)\frac{F_\Sun\cos B_\Sun}{F_{G}}\right]
	\sin(\varpi-\lambda_\Sun),
\end{equation}
\begin{equation}
\left<\frac{d\varpi}{dt}\right> =\frac{n}{e}\left[ \frac{3}{2}f(\epsilon)\frac{F_\Sun\cos B_\Sun}{F_{G}}\right]
	\cos(\varpi-\lambda_\Sun)+\dot{\varpi}_o,
\end{equation}
\begin{equation}
\left<\frac{di}{dt}\right> = -n\left[g(\epsilon)\frac{F_\Sun\sin B_\Sun}{F_{G}}\right]
	\cos(\Omega-\lambda_\Sun),
\end{equation}
\begin{equation}
\left<\frac{d\Omega}{dt}\right> = \frac{n}{\sin i}\left[g(\epsilon)\frac{F_\Sun\sin B_\Sun}{F_{G}}\right]
	\sin(\Omega-\lambda_\Sun)+\dot{\Omega}_o,
\end{equation}
where $\dot{\varpi}_o$ and $\dot{\Omega}_o$ are pericenter precession and nodal 
regression rates due to Saturn's oblateness.

Finally, we can simplify these expressions by replacing
the inertial longitudes $\varpi$ and $\Omega$ with 
longitudes measured relative to the Sun, 
$\varpi'=\varpi-\lambda_\Sun$ and 
$\Omega'=\Omega-\lambda_\Sun$:
\begin{equation}
\left<\frac{da}{dt}\right> = nae\left[2d(\epsilon)\frac{F_\Sun\cos B_\Sun}{F_{G}}\right]
	\sin\varpi',
\end{equation}
\begin{equation}
\left<\frac{de}{dt}\right> = n\left[\frac{3}{2}f(\epsilon)\frac{F_\Sun\cos B_\Sun}{F_{G}}\right]
	\sin\varpi',
\end{equation}
\begin{equation}
\left<\frac{d\varpi'}{dt}\right> = \frac{n}{e}\left[\frac{3}{2}f(\epsilon)\frac{F_\Sun\cos B_\Sun}{F_{G}}\right]
	\cos\varpi'+\dot{\varpi}'_o,
\end{equation}
\begin{equation}
\left<\frac{di}{dt}\right> = -n\left[g(\epsilon)\frac{F_\Sun\sin B_\Sun}{F_{G}}\right]
	\cos\Omega',
\end{equation}
\begin{equation}
\left<\frac{d\Omega'}{dt}\right> = \frac{n}{\sin i}\left[g(\epsilon) \frac{F_\Sun\sin B_\Sun}{F_{G}}\right]
	\sin\Omega'+\dot{\Omega}'_o,
	\label{dOdtav}
\end{equation}
where $\dot{\varpi}'_o=\dot{\varpi}_o-\dot{\lambda}_\Sun$ and
$\dot{\Omega}'_o=\dot{\Omega}_o-\dot{\lambda}_\Sun$ will be referred
to here as the ``modified" pericenter precession and nodal regression
rates, respectively.

Assuming that $B_\Sun$ changes sufficiently slowly, then for any semi-major 
axis $a$ there is a unique steady-state solution to these equations
where  $\left< \frac{da}{dt}\right>=\left< \frac{de}{dt}\right> = \left< \frac{d\varpi'}{dt}\right> = 
\left< \frac{di}{dt}\right> = \left< \frac{d\Omega'}{dt}\right> =0$.
This steady-state orbital solution has the following orbital parameters
(assuming $\sin i \simeq i$):
\begin{equation}
e_f=\frac{n}{\dot{\varpi}'_o}\left[\frac{3}{2}f(\epsilon)\frac{F_\Sun}{F_G}\cos B_\Sun\right],
\label{efeq}
\end{equation} 
\begin{equation}
\varpi_f=\lambda_\Sun+\pi,
\end{equation}
\begin{equation}
i_f=\frac{n}{|\dot{\Omega'}_o|}\left[g(\epsilon)\frac{F_\Sun}{F_G}\sin |B_\Sun|\right],
\label{ifeq}
\end{equation} 
\begin{equation}
\Omega_f=\lambda_\Sun+\frac{\pi}{2}\frac{B_\Sun}{|B_\Sun|}.
\end{equation}
This orbit has a finite eccentricity $e_f$ with the pericenter anti-aligned
with the Sun, so that the apoapse of the orbit points towards the Sun.
This is grossly consistent with the observed heliotropic behavior of the
Charming Ringlet shown in Figs.~\ref{longscans} and~\ref{elscan}. 
Furthermore, if 
$B_\Sun$ is non-zero, then this orbit also has a finite inclination, and
the ascending node is located $\pm90^\circ$ from the sub-solar longitude, 
depending on whether the Sun is north or south of the ringplane. 
The orbit will therefore be inclined so that it is on the opposite side of
the equator plane as the Sun at longitudes near local noon. 

\begin{figure}[tbp]
\resizebox{6in}{!}{\includegraphics{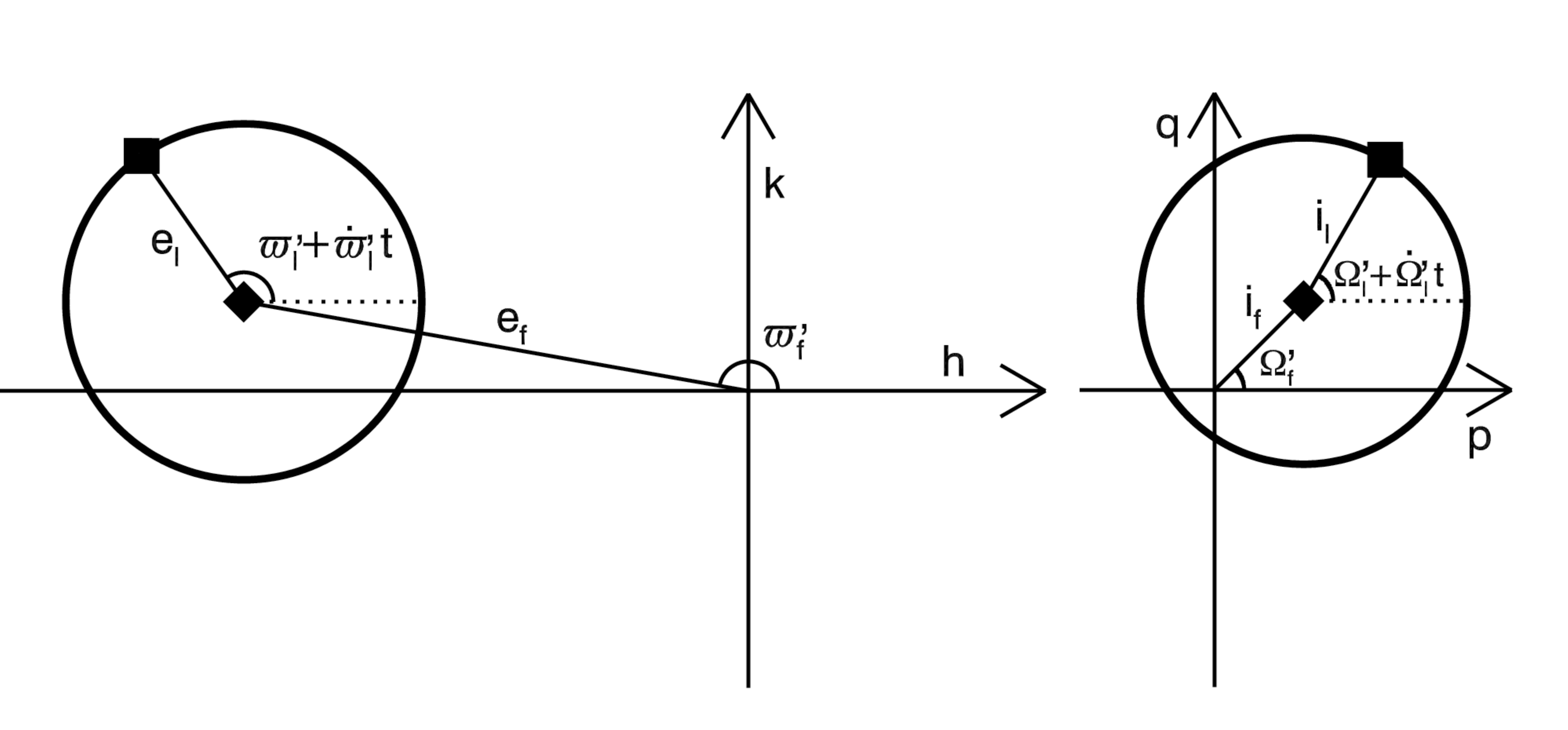}}
\caption{A graphical representation of the free and forced components 
of the eccentricity (left) and inclination (right), showing the parameters used in 
this analysis}
\label{diagpar}
\end{figure}

However, this steady-state solution is a special case.  More general solutions 
to the equation of motion can be most clearly described using the 
variables (Hor\'anyi and Burns 1991, see also Murray and Dermott 1999, equations 7.18-7.19):
\begin{equation}
h=e\cos(\varpi-\lambda_\Sun)=e\cos\varpi',
\end{equation}
\begin{equation}
k=e\sin(\varpi-\lambda_\Sun)=e\sin\varpi',
\end{equation}
\begin{equation}
p=i\cos(\Omega-\lambda_\Sun)=i\cos\Omega',
\end{equation}
\begin{equation}
q=i\sin(\Omega-\lambda_\Sun)=i\sin\Omega'.
\end{equation}
In terms of these variables, the above equations of motion reduce to:
\begin{equation}
\left<\frac{da}{dt}\right>=\frac{4d(\epsilon)}{3f(\epsilon)} a e_f \dot{\varpi}'_o k,
\label{aeqm}
\end{equation}
\begin{equation}
\left<\frac{dh}{dt}\right>=-\dot{\varpi}'_ok,
\label{heqm}
\end{equation}
\begin{equation}
\left<\frac{dk}{dt}\right>=\dot{\varpi}'_o(h+e_f),
\end{equation}
\begin{equation}
\left<\frac{dp}{dt}\right>=-\dot{\Omega}'_o\left(q-i_f\frac{B_\Sun}{|B_\Sun|}\right),
\end{equation}
\begin{equation}
\left<\frac{dq}{dt}\right>=\dot{\Omega}'_op,
\label{qeqm}
\end{equation}
where $e_f$ and $i_f$ are the steady-state (forced) eccentricity and inclination derived above.
While both these parameters depend on the semi-major axis $a$ via the mean-motion $n$, 
Eqn~\ref{aeqm} demonstrates that the fractional variations in the semi major axis are 
$\mathcal{O}(e^2)$ and can therefore be neglected for the nearly circular orbits of interest here.
Thus for the rest of this analysis $e_f$ and $i_f$ will be assumed to be constants.
In that case, $[h,k]$ and $[p,q]$ satisfy two pairs of separately coupled harmonic-oscillator equations, so
the trajectories traced out by the above equations form circles in $[h,k]$ and $[p,q]$ space \citep{HB91}. 
The centers of these circles are given by the steady-state solutions, and the
orbit evolves around the circles at rates given by the modified precession rates 
$\dot{\varpi}'_o$ and $\dot{\Omega}'_o$. As the orbit evolves along these paths, if $e_f$ and 
$i_f$ are nonzero, the orbit's total eccentricity and inclination will change
periodically with periods of $2\pi/\dot{\varpi}'_o$ and $2\pi/\dot{\Omega}'_o$, respectively.
In general, the various orbital parameters can be described as the vector sums
in $[h,k]$ and $[p,q]$ space of two components: a constant, ``forced" component and a 
time-variable, or ``free" component. The evolution of such an orbit is specified by 10 parameters
(see Fig~\ref{diagpar}):
\begin{itemize}
\item $e_f$ and $i_f$, the so-called forced eccentricity and forced inclination, whose
	values should be determined by Saturn's gravity and the in-plane and normal components 
	of the solar radiation pressure.
\item	$\varpi'_f=\varpi_f-\lambda_\Sun$ and $\Omega'_f=\Omega_f-\lambda_\Sun$,
	which specify the orientation of the orbit relative to the Sun. Given the above analysis, 
	we expect $\varpi'_f=\pi$ and $\Omega'_f=\pm\pi/2$	
\item $e_l$ and $i_l$, the so-called free eccentricity and free inclination. These
	are the radii of the circles traced out by the orbits in $[h,k]$ and $[p,q]$ space,
	respectively. These parameters can in principle have any non-negative value, and 
	are set by the initial conditions.
\item $\varpi'_l$ and $\Omega'_l$, the pericenter and node (relative to the Sun)
	of the free components of the eccentricity and inclination at some epoch time. 
	These parameters are initial conditions and can in principle have any value between
	0 and $2\pi$.
\item $\dot{\varpi}'_l$ and $\dot{\Omega}'_l$, which specify how fast 
	the free components of the eccentricity and inclination move
	around the fixed points. These parameters should equal
	the modified precession and regression rates 
	$\dot{\varpi}'_o$ and $\dot{\Omega}'_o$, which are determined by the 
	oblateness of the planet and the motion of the Sun.
\end{itemize}

At this point, it is useful to examine heuristically the evolution of such 
orbits, to clarify the physical processes involved. In the case where the
eccentricity equals $e_f$ and has its pericenter anti-aligned with the sun, 
the orbit can be regarded as a circular path that is displaced by a distance
$ae_f$ from Saturn's center. In this particular configuration, the orbit-averaged
torque on the particle from solar radiation pressure balances that
from the central planet, so the orbit does not evolve.
In other configurations, these torques will not balance and
the  eccentricity and pericenter will change over time. For example,
imagine that the orbit starts off with a small 
eccentricity $e_i<e_f$ and the pericenter located +90$^\circ$ ahead of the 
subsolar longitude. At this time, the particle is heading away from the Sun at
pericenter and is approaching the Sun at apocenter. The solar radiation 
pressure therefore causes the particle to accelerate in the direction of orbital 
motion when it is at pericenter and to decelerate at apocenter. This causes the orbit's
eccentricity  to increase
(note that in Eq.~\ref{dedt}, $F_t \cos f$ is positive in both positions). At the same time the pericenter precesses
around the planet under the influence of Saturn's oblateness. The eccentricity 
continues to grow until the apocenter becomes aligned with the Sun. However, 
once the orbit's precession carries the apocenter further, toward the dusk ansa of the ring, the
particle will be moving away from the Sun at apoapse and towards the Sun at periapse.
At this point, the solar radiation pressure will accelerate the particle near apoapse
and decelerate it near periapse, causing the eccentricity to shrink
(note that in Eq.~\ref{dedt}, $F_t \cos f$ is negative). The orbital eccentricity 
will therefore decrease until it reaches a minimum when the pericenter is 
aligned with the Sun, at which point the cycle begins anew. 

Now consider the inclination and node. Consider a case where
 the Sun is in the southern hemisphere, the initial inclination $i_i>i_f$,
and the ascending node is near the sub-solar longitude. At this time, the particle
is heading northwards on the sunward side of its orbit and southwards on the
shadowed side. The radiation force from the Sun pushes northwards 
on the particle as it passes on the sunward side of the planet, accelerating the vertical
motion of the particle and increasing the inclination in the orbit. If there were no shadow,
then this increased tilt would be cancelled out when the particle feels the same 
northward force as it is heading southward on the planet's far side. However, 
because sunlight is blocked from this side of the rings by the planet's shadow, the torque is not cancelled and the inclination increases. Meanwhile, the node regresses due to 
Saturn's oblateness (if $i>i_f$, then the second term on the left hand side of Eq.~\ref{dOdtav}
dominates). Thus the inclination continues to grow until the ascending node
reaches a point $90^\circ$ behind the Sun. After this point, the ascending node
will head into the shadow and the descending node will move towards the sub-solar
longitude. In this case, the particle is moving southwards while it is exposed to
solar radiation pressure that drives it northwards, so the radiation pressure will decelerate
the vertical motion and cause the inclination to lessen until it reaches a minimum
when the ascending node is $90^\circ$ ahead of the solar point, at which point
the cycle starts again. 

The orbital evolution described above is not specific to solar radiation
pressure, but will occur whenever the ring particles feel forces with
a fixed direction in inertial space. To demonstrate that
solar radiation pressure in particular is a reasonable explanation for the shape
and orientation of the Charming Ringlet, let us now  evaluate numerically the strength of the 
solar radiation pressure force $F_\Sun$ and the resulting $e_f$ and $i_f$. 

The solar radiation pressure  force $F_\Sun$ is given by \citep{BLS79}: 
\begin{equation}
F_\Sun=SAQ_{pr}/c,
\end{equation} 
where $c$ is the speed of light, $S$ is the solar energy  flux, 
$A$ is the cross-sectional area of the  particles, and  $Q_{pr}$ 
is an efficiency factor that is of order unity in the limit of geometric optics. 
The force ratio for quasi-spherical grains can therefore be written as:
\begin{equation}
\frac{F_\Sun}{F_G} = \frac{3}{4}\frac{S}{c}\frac{a^2}{GM}\frac{Q_{pr}}{\rho r_g},
\end{equation}
where $\rho$ is the particle's density and $r_g$ is the particle's radius.
 If we now assume $S=14$ W/m$^2$ at Saturn, $c=3*10^8$ m/s, 
$GM=3.8*10^{16}$ m$^3$/s$^2$, $a=$119,940 km (appropriate
for the Charming Ringlet) and $\rho=10^3$ kg/m$^3$ (appropriate
for ice-rich grains) we find:
\begin{equation}
\frac{F_\Sun}{F_G}=1.3*10^{-5}\frac{Q_{pr}}{r_g/1\hspace{2pt} \mu m}.
\label{forceratio}
\end{equation}
The other parameters in Eqs~\ref{efeq} and~\ref{ifeq}
can also be estimated. For the observations considered here, 
the shadow covers roughly 80$^\circ$ in longitude, so $\epsilon \simeq 0.2$, 
in which case  $f(\epsilon) \simeq 0.85$ and $g(\epsilon) \simeq 0.2$. 
Also,  given Saturn's gravitational harmonics 
\citep{Jacobson06},  the orbital and precession
rates in the vicinity of the Charming Ringlet are $n=736^\circ$/day and 
$\dot{\varpi}'_o\simeq|\dot{\Omega}'_o|\simeq 4.7^\circ/$day.
With these values, the forced eccentricities and inclinations are: 
\begin{equation}
e_f\simeq 0.0026\cos B_\Sun\frac{Q_{pr}}{r_g/1\hspace{2pt}  \mu m},
\label{epred}
\end{equation}
\begin{equation}
i_f\simeq 0.00041\sin| B_\Sun|\frac{Q_{pr}}{r_g/1\hspace{2pt} \mu m}.
\label{ipred}
\end{equation}
Note that over the course of Saturn's year, $\cos B_\Sun$
ranges from 0.9 to 1.0, while $\sin|B_\Sun|$ ranges from 0 to
0.5. Thus $i_f$ can change significantly on seasonal time scales, 
while $e_f$ should remain approximately constant.

The $ae$ observed in the Charming Ringlet range
between 10 and 30 km. This would be consistent with the 
$e_f$ predicted by this model if $r_g/Q_{pr}$ is between 
10 and 30 microns, which are perfectly reasonable values.  
These findings therefore support the notion that solar radiation 
pressure influences  this ringlet's dynamics. 

The variations in the ringlet's eccentricity, pericenter and node
relative to the Sun could potentially also be explained by
this sort of model in terms of non-zero free eccentricities
and inclinations. Indeed, we will show below that just such a model
can provide a useful description of the ring's shape. However, 
at the same time, we must recall that the above analysis
was for the orbital properties of a single particle, whereas 
the observed ringlet is composed of many particles. One would expect that
these particles would have a range of sizes, and  some dispersion in their 
orbital parameters. While the shape of the ringlet should reflect
the average orbital parameters of all its constituent particles, 
one might have expected that this averaging would wash out
any free component in the eccentricity or inclination. 
Such a model therefore raises a number of questions
about the dynamics of this ringlet, which will be discussed
in more detail below.

\section{Combining the observations}

\begin{figure}
\resizebox{6in}{!}{\includegraphics{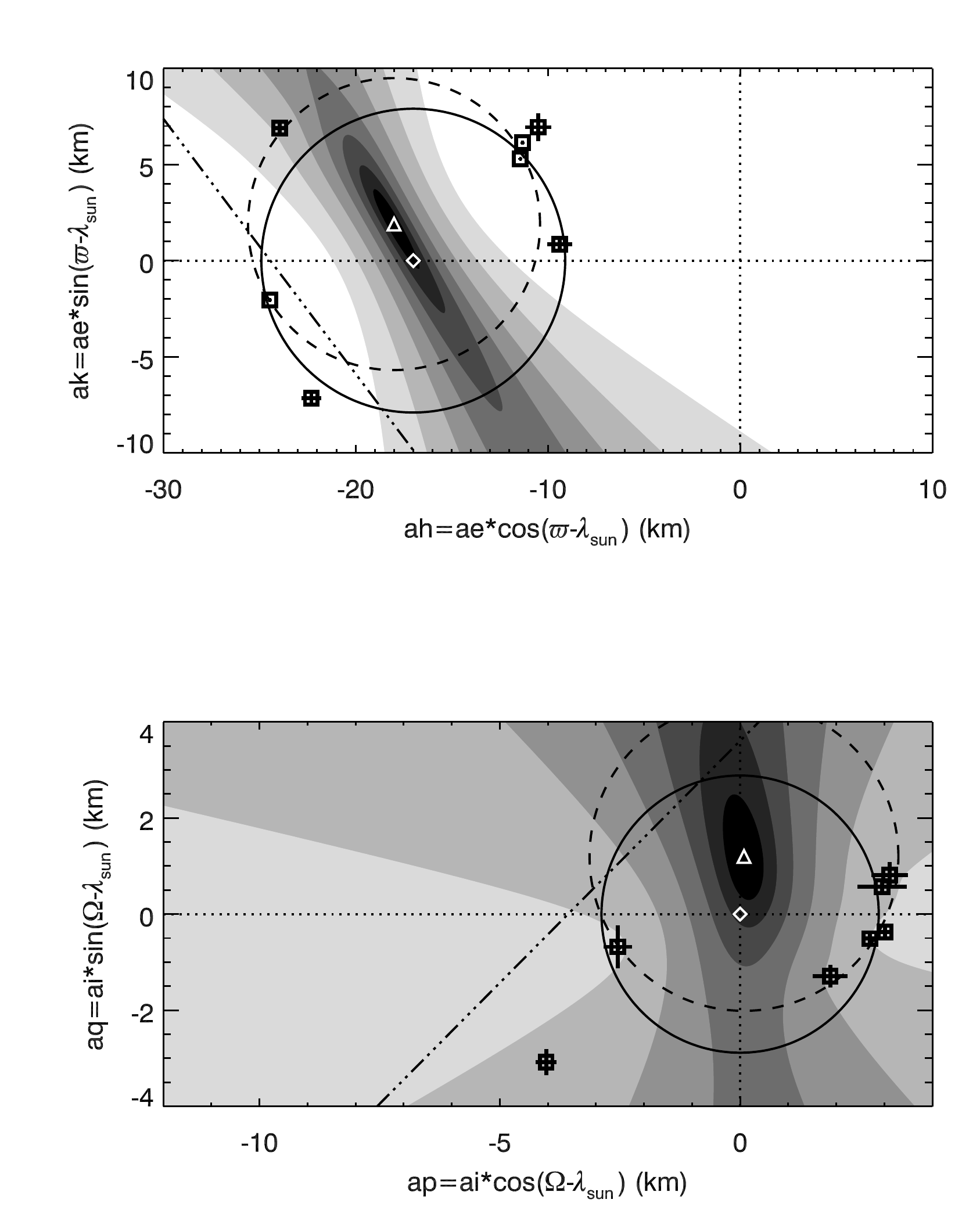}}
\caption{Diagrams showing the data derived from the various observations
versus the parameters $ah$ and $ak$ (top) and $ap$ and $aq$ (bottom). The data points
with error bars come from the longitudinal scans (see Table~\ref{longpartab}). The 
dot-dashed line represents the constraints from the elevation scan. The gray scales
in the background indicate the values of the $\chi^2$-statistic on the forced
eccentricity and inclination discussed in the text. Each level corresponds to a factor
of 2 in $\chi^2$ (which dark being lower values). The white triangles mark the best-fit values of $h,k,p$ and $q$.
In the upper plot, the white diamond marks the best-fit forced eccentricity along the $k=0$
axis, while in the lower plot the diamond is at the origin. The solid and dashed lines
are the best-fit circles centered on the diamonds and triangles, respectively.}
\label{charmpar}
\end{figure}

Keeping in mind the above caveats about applying a model appropriate to an individual
particle's orbit to the entire ringlet, we will now attempt
to fit the observational data to a ten-parameter global model that includes
both forced and free components in the eccentricity and inclination.

As discussed above, an orbit with forced and free orbital elements will trace out circles
in $[h,k]$ and $[p,q]$ space as the orbit evolves. Therefore, we plot the orbital elements
derived from the above fits in this space (see Fig.~\ref{charmpar}). 
Intriguingly, the admittedly  sparse data do seem to 
describe a circle in $[h,k]$ space, centered roughly at $[ah,ak]=[-17,0]$ km. In $[p,q]$ space, the
situation is less clear. Neglecting the outlying data from Orbit 96 (discussed above), the data could be 
consistent with a circle centered near the origin, but most of the data points
are clustered to one side of the circle, making it difficult to be certain.

To make these visual impressions more quantitative, we found the circles in $[h,k]$ and $[p,q]$
that best describe the data. We used the following procedures 
for the $[h,k]$ data: For each possible value of $[h,k]$, we computed
the distance between this point and the location of every one of the longitudinal scan data points
[$h_j$, $k_j$]:
\begin{equation}
R_j(h,k)=\sqrt{(h-h_j)^2+(k-k_j)^2}.
\label{req}
\end{equation}
(Note the data from the elevation scan are not included in this analysis because
they do not constrain $h$ and $k$ or $p$ and $q$ separately).
We also calculate a typical error for each data point $\sigma_j$, which is the average
of the errors  on $h$ and $k$ (the difference in the errors on these parameters 
was not considered large enough to justify complicating the analysis). 
We then compute the average value of the 
appropriate $R_j$, weighting the observations by their error bars, to obtain the mean 
distance $\bar{R}$. We then compute the following $\chi^2$ statistic:
\begin{equation}
\chi(h,k)^2=\sum \frac{(R_j(h,k)-\bar{R})^2}{\sigma_j^2}.
\end{equation}
This statistic measures the goodness of fit of the data to the best-fit circle centered at a given
value of $h$ and $k$. The $[p,q]$ space analysis is essentially
the same, except that the data from Orbit 96 are excluded from the fit for
the reasons described above. The contours in Fig.~\ref{charmpar} illustrate 
how $\chi^2$ varies with $[h,k]$ and $[p,q]$.

For the $[h,k]$ plot, the best-fit solution is at $[ah,ak] = [ -18,1.9]$ km. This would imply that the periapse
leads the anti-solar direction by 6$^\circ$. However, the best-fit solution assuming the pericenter
is exactly anti-aligned with the Sun is not obviously worse than the overall best fit.  
Note that even for these best-fitting models, the $\chi^2$ fit is still quite poor 
(74 for 4 degrees of freedom). This is consistent with a visual inspection of the
data, which scatter around the circle by more than their error bars. This excess scatter
could occur for a number of reasons. The data used here come from a range of phase angles
and are sensitive to different parts of the size distribution, which may lead to differences in the apparent shape of the ringlet. Also, our background subtraction algorithm and other
procedures used to derive the radial positions of the ringlet may have introduced
systematic errors between different scans. 

For the $[p,q]$ plot, the best-fitting model has $[ap,aq]=[0,1.2]$ km.  Here the $\chi^2$ value
is good (3.1 for 3 degrees of freedom). However, the difference
in the quality of the fit between this and $[p,q]=[0,0]$ is only marginally
significant (assuming no forced inclination, the $\chi^2$ is 9.9 for 5 degrees of freedom). 
Furthermore, since the Sun is in the southern hemisphere, and $B_\Sun$ is negative, 
we expect that the best-fit $q$ should be negative, not positive. Thus the best-fitting model is a bit of
a surprise.  Since $B_\Sun$ changes significantly over the time period covered
by these observations, a more complete model would include a time-variable forced inclination.
However, given the weak evidence for any forced inclination at all, we chose not to
consider such complications at this time.

Despite these uncertainties, we can now explore whether the temporal evolution of the shape parameters are consistent with the above model, which suggests that the parameters
should drift around the circles at nearly constant rates determined by the 
modified pericenter-precession and nodal-regression rates. Given the sparseness
of the data, we cannot establish easily whether any given solution is unique. However, preliminary
examination of the data showed that they were approximately consistent with
the expected drift rates ($\dot{\varpi}'_o\simeq-\dot{\Omega}'_o\simeq4.7^\circ$/day).
Therefore, for each posssible solution for $e_f$, $e_l$, $i_f$ and $i_l$, we
determined the phase of the shape for each longitudinal-scan observation, unwrapped
the phase assuming drift rates close to those expected, and fitted the resulting
phases versus observation time to a line to obtain estimates
of the rates $\dot{\varpi}'_{l}$ and $\dot{\Omega}'_{l}$, as well as the
longitudes at epoch $\varpi'_{l}$  and $\Omega'_{l}$ (the epoch time being taken
as the time of the first image in the Orbit 42 sequence 2007-099T22:19:10, see Table~\ref{obstab1}).
Regardless of whether we accepted the best-fit solution
(triangles/dashed circles in Fig~\ref{charmpar}) or a simplified solution
assuming $\varpi'_f=180^\circ$ and $i_f=0$ (diamonds/solid circles in Fig~\ref{charmpar}),
we obtain roughly the same rates. $\dot{\varpi}'_l=4.66^\circ$/day and 
$\dot{\Omega}'_l=-4.75^\circ$/day. Recall that these are the modified 
precession rates in a reference frame tied to the Sun. The precession rates
in an inertial coordinate system must account for the movement of Sun 
$\dot{\lambda}_\Sun=0.03^\circ$/day. Thus the precession rates are actually:
$\dot{\varpi}_l=4.69^\circ$/day and $\dot{\Omega}_l=-4.72^\circ$/day.
The expected rates at 119940 km are 4.71$^\circ$/day and -4.68$^\circ$/day, 
respectively, so these numbers are close to theoretical expectations. This  
model therefore can provide a useful parametrization of the
available data.

Table~\ref{modparams} summarizes the model parameters for the shape
of the Charming Ringlet. Model 1 is the simplified model in which $\varpi_f$
is taken to be exactly anti-aligned with the Sun and the forced inclination
is assumed to be zero. Model 2 is the more complex model that allows both
$\varpi'_f$ and $i_f$ to have their ``best-fit" values.

\begin{table}
\caption{Model parameters for the Charming Ringlet}
\label{modparams}
\resizebox{6in}{!}{\begin{tabular}{|c|c|c|c|c|c|c|c|c|c|c|}\hline
Model &  $ae_{f}$  & $\varpi'_f$ & $ae_{l}$ $^a$ &$\dot{\varpi}'_{l}$ $^b$ & $\varpi'_{l}$ $^{b,c}$ 
	& $ai_{f}$ & $\Omega'_f$ &  $ai_{l}$ $^a$ & $\dot{\Omega}'_{l}$ $^b$ & $\Omega'_{l}$ $^{b,c}$ \\
& (km) & (deg) & (km) & (deg/day) & (deg) & (km) & (deg) & (km) & (deg/day) & (deg) \\
	\hline 
1 & 17.0$\pm$0.5$^d$ & 180 & 7.9$\pm$0.4 & 4.66$\pm$0.01 & 230$\pm$3 
	& -- & -- &2.9$\pm$0.2 & -4.73$\pm$0.02 &  -152$\pm$9 \\
2 & 18.1 & 174 &  7.6$\pm$0.4 & 4.67$\pm$0.01 & 225$\pm$4
	& 1.3 & +90 & 3.3$\pm$0.1 & -4.77$\pm$0.02 & -158$\pm$9 \\
\hline
\end{tabular}}

$^a$ Errors are the standard deviations of the values $R_j$, see Eqn~\ref{req}.

$^b$ Errors from linear fit, assuming central values for $ae_f, ae_l, ai_f$ and $ai_l$

$^c$ Longitudes relative to Sun at epoch=2007-099T22:19:10 (time of first observation in Orbit  042).

$^d$ Error based on factor of 2 increase in $\chi^2$ relative to best-fit value 

Note Rev 96 data excluded from inclination/node fits.
\end{table}

\begin{table}[t]
\caption{Comparison of model predictions for eccentricity and pericenter with 
observed longitudinal scan data}
\label{modcompe}
\resizebox{6in}{!}{\begin{tabular}{|l|c|c|c|c|c|c|c|c|c|c|}\hline
Orbit/Obs. Sequence & ae (km) & ae (km) & ae (km) & Observed- & Observed- & 
	$\varpi'$ (deg) & $\varpi'$ (deg) & $\varpi'$ (deg) & Observed- & Observed-  \\
& Observed & Model 1 & Model 2 & Model 1 & Model 2 &
	Observed & Model 1 & Model 2 & Model 1 & Model 2 \\\hline
030/AZDKMRHP001/ & 
24.9 & 23.3 & 23.5 & +1.6 & +0.8 & 163.9 & 166.1 & 161.1 & -2.1 & +2.9 \\
042/RETMDRESA001/ & 
23.4 & 22.9 & 23.9 & +0.5 & -0.2 & 197.8 & 195.3 & 188.6 & +2.5 & +9.2 \\
070/RETMDRESA001/ & 
 9.4 &  9.4 & 10.7 & +0.0 & -1.7 & 174.8 & 170.0 & 161.9 & +4.9 & +12.9 \\
071/PAZSCN002/          & 
12.5 & 13.2 & 14.1 & -0.6 & -2.4 & 146.5 & 153.3 & 149.7 & -6.8 & -3.2 \\
082/RETARMRLP001/ & 
12.9 & 13.8 & 14.9 & -0.9 & -2.8 & 151.6 & 152.7 & 149.3 & -1.1 & +2.3 \\
092/RETARMRLF001/ & 
12.6 & 12.2 & 13.4 & +0.4 & -1.6 & 155.2 & 154.9 & 150.5 & +0.2 & +2.6 \\
 096/RETARMRMP001/  & 
24.6 & 24.9 & 25.5 & -0.3 & -1.0 & 184.8 & 182.2 & 178.3 & +2.6 & +6.5 \\ \hline
Mean & & & & +0.11 & -1.27 & & & & +0.02 & +5.04 \\
St. Dev. & & & & 0.86 & 1.29 & & & & 3.85 & 5.19 \\ 
\hline
\end{tabular}}
\end{table}

\begin{table}
\caption{Comparison of model predictions for inclination and node with
observed longitudinal scan data}
\label{modcompi}
\resizebox{6in}{!}{\begin{tabular}{|l|c|c|c|c|c|c|c|c|c|c|}\hline
Orbit/Obs. Sequence & ai (km) & ai (km) & ai (km) & Observed- & Observed- & 
	$\Omega'$ (deg) & $\Omega'$ (deg) & $\Omega'$ (deg) & Observed- & Observed-  \\
& Observed & Model 1 & Model 2 & Model 1 & Model 2 &
	Observed & Model 1 & Model 2 & Model 1 & Model 2 \\\hline
030/AZDKMRHP001/ & 
2.3 & 2.9 & 2.6 & -0.6 & -0.3 & 325.3 & 316.1 & 337.5 & +9.3 & -12.7 \\
042/RETMDRESA001/ & 
2.6 & 2.9 & 3.1 & -0.3 & -0.4 & 195.0 & 207.7 & 179.2 & -12.7 & +15.8 \\
070/RETMDRESA001/ & 
3.2 & 2.9 & 3.9 & +0.3 & -0.6 & 14.6 & 33.1 & 33.1 & -18.5 & -18.5 \\
071/PAZSCN002/          & 
3.0 & 2.9 & 3.1 & +0.1 & -0.1 & 10.9 & -4.0 & -0.4 & +14.8 & +11.2 \\
082/RETARMRLP001/ & 
3.0 & 2.9 & 2.8 & +0.1 & +0.3 & 352.9 & 345.0 & 345.3 & +7.9 & +7.6 \\
092/RETARMRLF001/ & 
2.8 & 2.9 & 2.8 & +0.1 & -0.1 & 349.3 & 351.9 & 350.2 & -2.5 & -0.8 \\
 096/RETARMRMP001/  & 
5.1 & 2.9 & 3.6 & +2.2 & +1.5 & 217.3 & 203.3 & 156.8 & +14.1 & +60.5 \\ \hline
Mean (excluding 096 data) & & & & -0.06 & -0.21 & & & & -0.28 & +0.53 \\
St. Dev. (excluding 096 data) & & & & 0.34 & 0.33 & & & & 13.24 & 13.54 \\ 
\hline
\end{tabular}}
\end{table}

\begin{table}
\caption{Comparison of model predictions with elevation scan observation}
\label{modcompel}
{\begin{tabular}{|l|c|c|c|c|c|c|} \hline
 & $ae$ (km) & $\varpi'$ (deg) & $ai$ (km) & $\Omega'$ (deg) & $C$ (km/rad) & $z$ \\ \hline
 Observation & & & & & 19.5 & 2.54 \\
 Model 1 & 24.3 & 188.2 & 2.9 & 237.7 & 22.0 & 2.78 \\
 Model 2 & 25.2 & 186.2 & 2.4 & 216.6 & 22.4 & 2.37 \\ \hline
 \end{tabular}}
 \end{table}

\clearpage

\section{Comparing model predictions with the observations}

 Tables~\ref{modcompe}, \ref{modcompi} and~\ref{modcompel} compare the observed
 shape parameters measured by the various observations with the predictions from the
 two models derived above. While the model parameters were derived using weighted
 averages of data from different observations, these comparisons do not consider
 variations in the uncertainties in the observations. This is because, as noted above, 
 these simplified models were unable to fit the $[h,k]$ data to within the error bars. Thus
 an unweighted analysis will provide a conservative estimate of how well these models
 describe the data.
 
 Table~\ref{modcompe} presents the model predictions for the eccentricity and pericenter
 locations from the longitudinal scans. Note that the only observation where the more 
 complex Model 2 does a better
 job predicting the eccentricity and pericenter than the simpler Model 1 is in the Orbit 30 data.
 This is consistent with Fig~\ref{charmpar}, where the dashed circle (Model 2) gets closer
 to the point in the upper left (from Orbit 30) than the solid circle (Model 1), but for all
 the other data points the dashed circle is not obviously a better fit than the solid one.
Note the Orbit 30 data were taken at a substantially higher phase angle than the other observations,
so this observation may probe a different part of the size distribution
and the shape parameters may not be perfectly comparable to the others.  
Therefore, we conclude that the simpler
model that assumes the forced component of the pericenter is perfectly anti-aligned with the Sun is a preferable
model for the shape of the ring. This model recovers the eccentricity of the ringlet
with an $rms$ residual of 1 km and the pericenter location with an $rms$ residual of 4$^\circ$.

Table~\ref{modcompi} presents the model predictions for the inclinations and nodes for the
longitudinal scans. 
In this case, there is not a clear difference between the  two models. Given that including
a forced inclination does not substantially reduce the scatter in the observations, for 
the sake of simplicity we favor the use of the simpler Model 1 in this 
case as well. Here the model predicts the inclination with an $rms$ residual of 0.3 km and the node location with an $rms$ residual of 14$^\circ$. 

Finally, Table~\ref{modcompel} compares the model predictions for the $z$ and $C$ parameters
for the elevation scan (see Eqs~\ref{zeq} and~\ref{ceq}). This is a critical check on the model, which was developed using only
the longitudinal scan data. Here, we can see that both models give values for $z$ and $C$ that
are reasonably consistent with the observed values. 

In conclusion, while Model 1 is clearly over-simplified and does not provide a perfectly
accurate description of the observed data, it nevertheless appears to be a useful approximate
description of the ringlet's  shape and time variability. 

\section{Interpretation}

We can now compare
the observed shape parameters of this ringlet with theoretical expectations. The {\sl forced}
eccentricity and inclination can be relatively easily understood
in terms of the solar radiation forces discussed above. By contrast, 
the {\sl free} components of the eccentricity and inclination are surprising and more difficult 
to explain.

\subsection{Forced eccentricity and inclination}

Equations~\ref{epred} and~\ref{ipred} indicate that  solar radiation
pressure should produce a ring with $i_f/e_f \simeq 0.16\tan|B_\sun|$. 
For the observations described here $|B_\Sun|$ ranges between 16$^\circ$ and $3^\circ$,
so $i_f/e_f$ would be between 0.04 and 0.01. This is consistent with the observed
values of $ae_f \simeq 17$ km and $ai_f<1$ km.  Furthermore, the observed
$ae_f$ suggests a typical particle size $r_g \simeq 20 \mu$m$*Q_{pr}$, which is not unreasonable. However, we must caution that 
the particles in the Charming Ringlet probably have a distribution of  sizes, and 
this estimated value of $r_g$ may only be an effective average value for this distribution.
The particle size distribution will be investigated in more detail
in a future study of the ring's  spectrophotometric
properties and detailed morphology.

\subsection{Free eccentricity and inclination}

While nonzero free eccentricities and inclinations are acceptable solutions to the equation of 
motion for a single particle's orbit, it is surprising for the ringlet as a whole to 
exhibit such terms, because they imply that all the component particles' orbits 
not only have comparable finite values of $e_l$ and $i_l$, but also have
similar values of $\varpi_l$ and $\Omega_l$. Such an asymmetry in these components of the ring's shape could be due to one of three things: (1) an asymmetry in the initial conditions of the ring particles, (2) an explicit longitudinal asymmetry in the equations of motion, or (3) a spontaneous symmetry-breaking in the ringlet. We will consider each of the possibilities below. 

\subsubsection{Asymmetric initial conditions}

There are various ways to produce a collection of particles with the
same values for $\varpi_l$ and $\Omega_l$. For example,  
an impact near the present location of the ringlet could release a cloud of dust from one
point in space, suddenly injecting a collection of particles into the gap that have similar 
orbital elements. Alternatively, particles could be supplied into the
ring over an extended period of time, but for some reason dust grains with certain 
orbital parameters are generated at higher rates than others. In this case, the relevant
source bodies for the dust would almost certainly be too large to have any
detectable forced eccentricity due to solar radiation pressure. Thus
the observed heliotropic ring could not be simply be low-velocity impact debris tracing the orbit of its source material, but instead must reflect some more complex  production process
involving various interactions with the local plasma and dust environment.

Regardless of how the particles were injected into the ring, the observable ring particles
must be relatively young in order for any asymmetry in the initial conditions to be visible in
the present ringlet. The Charming Ringlet has a full-width at half-maximum of 
about 30 km. If we assume a comparable spread in semi-major axes, then the
precession rates of the particles in the ring will vary by about 0.003$^\circ$/day. 
The values of $\varpi_l$ and $\Omega_l$ would therefore spread over all possible
longitudes in  a few hundred years. While this time-scale could be extended if we 
assume the radial width of the ring is due to variable eccentricities rather than 
semi-major axes, even then the visible particles in the ring probably cannot be more than a 
few thousand years old if they are to preserve any asymmetry in their initial conditions.
Such ages are not entirely unreasonable, for small dust grains like those seen in the Charming Ringlet can be rapidly destroyed by energetic particle bombardment \citep{BHS}, or lost by adhering to larger objects
in the Cassini Division. However, we must caution that the production and loss of dust grains
within narrow gaps has not been studied in great detail yet.

Another important constraint on these sorts of models is the lack of gross variations
in the brightness or morphology of the ring with longitude. This argues against any
large source bodies existing within the ringlet itself, as such objects would tend to scatter
and perturb the material in their vicinity, producing either gaps or possibly clumps 
similar to those visible in the Encke gap ringlets; such gaps and clumps are 
not seen in the Charming Ringlet. It also requires that the ringlet grains exist long 
enough to spread evenly over all longitudes, which takes a few years or 
decades. 

\subsubsection{Asymmetric terms in the equations of motion}

Instead of an asymmetric source, it is also conceivable that
the equations of motion contain terms that depend on $\varpi_l$
and $\Omega_l$. Recently, \citet{Hedman10} demonstrated
that a combination of perturbations from Mimas and the massive B-ring outer edge could
give rise to terms in the equation of motion like:
\begin{equation}
\left<\frac{d^2\varpi}{dt^2}\right> = -f_o^2\sin(\varpi-\dot{\varpi}_r t),
\label{perilock}
\end{equation}
where $f_o$ and  $\dot{\varpi}_r$ are constants. Such a term
acts as a restoring force on the pericenter location of any particle's orbit.
Thus, in a region where the precession rate $\dot{\varpi} \simeq \dot{\varpi}_r$, 
this term aligns the pericenters of all freely-precessing eccentric orbits. If such
a term was effective on the Charming Ringlet, it could
explain how all the particles in the ringlet happen to have the same value of 
$\varpi_l$.

One difficulty with this sort of model is that the particles
in the Charming Ringlet seem to have both  $\varpi_l$ and $\Omega_l$ 
aligned. While one could imagine
expressions similar to Eqn~\ref{perilock} involving the node instead of 
the pericenter, it is difficult to have both terms operate at the same
location. Like any other resonant term in the equations of motion, such 
terms can only be effective over a narrow range of semi-major axes
(or equivalently, narrow ranges of  $\dot{\varpi}$ and/or $\dot{\Omega}$), 
and resonances involving nodes typically occur at different
locations from those involving pericenters \citep{MurrayDermott}. It therefore
would be quite a coincidence if the Charming Ringlet just
happened to fall at a location where both angles could be 
effectively constrained.

\subsubsection{Spontaneous symmetry breaking}

A ringlet with finite free eccentricity and free inclination can in principle 
form  spontaneously without any terms in the equations of motion that depend 
explicitly on $\varpi_l$ and $\Omega_l$, and without any
strong asymmetry in the particle's initial conditions. 
Such phenomena have been discussed almost exclusively
in the context of massive, dense ringlets \citep{BGT85}.
However, one can argue that this sort of
``spontaneous symmetry-breaking''  could also occur
in low-optical-depth dusty rings via dissipative processes 
like collisions, provided that 
there are terms in the individual particle's equations of motion that favor
the development of a nonzero $e_l$ and $i_l$ comparable 
to those observed for the entire ringlet. 

Dissipative collisions are often invoked as a 
mechanism that causes narrow rings to spread 
in semi-major axis \citep{GT82},
so it might seem surprising that such collisions
could also align pericenter or node locations. However,
unlike the semi-major axis, the longitudes of 
pericenter and node have no direct effect on a 
particles' orbital energy. Thus, while the dissipation
of orbital energy requires that particles' orbital semi-major axes
evolve in a particular direction, this is 
not the case for pericenters or nodes. Instead, the
evolution of pericenters and nodes should be
driven primarily by the collisons' dissipation of 
relative motions.

To illustrate how such collisions can align pericenters
and nodes, consider the following simple situation:
There is a ringlet composed of many particles
with similar orbital properties, and there is a single particle whose orbit
is misaligned with the others. For simplicity, assume that both
the ringlet and the particle have zero eccentricity and zero forced inclination. 
Furthermore, assume that both the ringlet and the particle have the same free inclination $i$ but
different longitudes of ascending node $\Omega_r$ and $\Omega_p$, respectively.
If $\Omega_r\ne \Omega_p$, then the particle's orbit will
cross the ringlet at two longitudes $\lambda_c=(\Omega_p+\Omega_r)/2 \pm \pi/2$.
At these two longitudes the particle will feel a force due to its collisions with the 
particles in the ringlet, and the vertical component of that force $F_z$ will
be proportional to the vertical velocity of the particles in the ringlet, so
$F_z \propto \cos(\lambda_c-\Omega_r)$. Inserting this into Equation~\ref{dOdt},
we can express the perturbation to the particle's node position due to its interactions
with the ringlet as:
\begin{equation}
\frac{d\Omega_p}{dt} = 2D\sin (\lambda_c-\Omega_p)\cos(\lambda_c-\Omega_r)
\end{equation}
where $D$ is a constant.
Substituting in the above expression for the crossing longitudes $\lambda_c$ and simplifying, 
this expression reduces to the simple form:
\begin{equation}
\frac{d\Omega_p}{dt} = -D\sin (\Omega_p-\Omega_r).
\end{equation}
The forces applied to the particle's orbit during the ringlet crossings therefore
do tend to align the particle's orbital node position with that of the ringlet.
A similar calculation demonstrates that the same basic phenomenon acts
to align pericenters as well. Thus collisions can indeed align 
pericenters and nodes, provided the
collisions are frequent (and lossy) enough, and provided the particles
maintain some finite (free) eccentricity and inclination. 

The requirement that collisions are frequent enough to align pericenters 
is probably met for the Charming Ringlet. While this ringlet has a 
low normal optical depth (roughly 10$^{-3}$), the orbital period is 
sufficiently short (around 0.5 days) that the collisional timescale is still 
only a few years or decades, much less than the typical erosion
timescales of thousands of years~\citep{BHS}. 

On the other hand, the persistance of the nonzero free eccentricities 
and inclinations probably requires some modifications to the individual 
particles' dynamics. If the particles'  equations of motion were just given by 
Eqs~\ref{heqm}-~\ref{qeqm} above, dissipative collisions would (assuming
the initial conditions were not highly asymmetric)
tend to produce a ringlet with $e_l=i_l=0$. Thus, we probably need
to add some additional terms to these equations to produce 
something similar to the Charming Ringlet's  observed shape. 
One relatively simple way to accomplish this is to 
add non-linear damping terms into the equations:
\begin{equation}
\left<\frac{dh}{dt}\right>=-\dot{\varpi}'_ok
	+\gamma_h (h+e_f)\left[1-\left(\frac{h+e_f}{e_l}\right)^2\right],
\end{equation}
\begin{equation}
\left<\frac{dk}{dt}\right>=\dot{\varpi}'_o(h+e_f)
	+\gamma_k k\left[1-\left(\frac{k}{e_l}\right)^2\right],
\end{equation}
\begin{equation}
\left<\frac{dp}{dt}\right>=-\dot{\Omega}'_o\left(q-i_f\frac{B_\Sun}{|B_\Sun|}\right)
+\gamma_p p\left[1-\left(\frac{p}{i_l}\right)^2\right],
\end{equation}
\begin{equation}
\left<\frac{dq}{dt}\right>=\dot{\Omega}'_op
	+\gamma_q \left(q- i_f\frac{B_\sun}{|B_\Sun|}\right)\left[1-\left(\frac{q - i_fB_\sun/|B_\Sun|}{i_l}\right)^2\right],
\end{equation}
where $\gamma_h, \gamma_k << \dot{\varpi}'_o$ and 
$|\gamma_p|, |\gamma_q| << |\dot{\Omega}'_o|$ quantify the magnitude of the damping 
terms. These terms transform the $[h,k]$ and $[p,q]$ systems 
from simple harmonic oscillators into van der Pol oscillators \citep{Baierlein}. 
Such oscillators are characterized by a limit cycle which
the system will asymtotically approach no matter where it is started
in $[h,k]$ and $[p,q]$ space. These limit cycles are circles centered 
at $[h,k]=[-e_f,0]$ and $[p,q]=[0,\pm i_f]$ with radii of $e_l$ and $i_l$, 
and the orbit  traces out the circles at rates given 
by $\dot{\varpi}'_o$ and $\dot{\Omega}'_o$. These equations of motion
therefore cause any particle's orbit to evolve to the same path as the 
observed ringlet. On their own, particles started at different points in phase space
will wind up at different points along this cycle. However, if
the relative motions among the particles are efficiently dissipated, then all the particles
should eventually clump together in phase space such that they all move around the
limit cycle together, as observed.

Models of this sort  have the advantage that the
additional terms in the equations of motion do not have explicit frequency-dependent
terms that can only be effective at specific locations in the rings. Such
terms are therefore more likely to show up in a broader range of contexts, and could
even be generic features of small dust grains' dynamics in narrow gaps. 
For example, the nonlinear damping terms in the above equations contain either
$e_l$ or $i_l$. While these are small numbers in absolute terms, they are
not much smaller than the fractional gap width $\delta a/a$, so such factors 
could arise due to interactions between the ringlet particles and the gap edges. 
This would not be unreasonable, as small particles could be
attracted to the edges by the force of gravity, or even repelled if the small grains
in the ring have a sufficient electrical charge. Furthermore, variations in the plasma environment 
within the gap could also possibly produce perturbations on the grains' motions 
with the appropriate positional dependence.

One clue to the exact nature of these forces is that the observed
ring traces out a circle that is centered on the point $[h,k]=[-e_f,0]$ and
excludes the origin $[h,k]=[0,0]$. Based on some preliminary analyses,
it appears that a limit cycle of this type cannot be created by non-linear
damping terms involving only  $e$ or $\varpi$ , but instead requires
terms that contain $k$ and/or $h+e_f$, like the ones given above  (note that 
only one of the two terms $\gamma_h$  and $\gamma_k$ has to be 
non-zero to produce the desired limit cycle). Since $h$ and $k$ are tied to
the location of the Sun, this implies that these damping terms might also
have some connection with the Sun. One possibility is that these terms
reflect the influence of Saturn's shadow. When small particles enter the shadow,
electrons are no longer being ejected from their surfaces via the photoelectric effect. 
This can significantly change their electric charge and thereby lead to significant
forces that would preferentially damp or drive $h$ or $k$. Further investigation
is needed to explore whether the perturbations from these or other processes 
could account for the observed shape of the Charming Ringlet.

\section*{Acknowledgments}

We acknowledge the support of NASA, the Cassini Project and the Imaging Science Team
for obtaining the images used in this analysis. This work was also supported by
Cassini Data Analysis Program grants NNX07AJ76G and NNX09AE74G. We wish to
thank P.D. Nicholson and M.R. Showalter for useful conversations, and S. Charnoz and
J. Schmidt for their helpful reviews of this manuscript.

\section*{Appendix: Orbit averages including shadow effects}

Equations~\ref{a1a}-\ref{o1a} are derived from Equations~\ref{a1}-\ref{o1}
by averaging over all longitudes $\lambda$. This averaging procedure
is complicated by the presence of Saturn's shadow, which blocks sunlight from 
reaching part of the rings. If a fraction $\epsilon$ of the ring is in Saturn's shadow,
then the ring particles only feel the solar radiation pressure when 
$|\lambda-\lambda_\Sun|<\pi(1-\epsilon)$. Thus if $X$ is 
any of the radiation-pressure-induced terms on the right-hand sides of  
Eqs~\ref{a1}-\ref{o1}, then the orbit-averaged value of $X$ is:
\begin{equation}
<X> =\frac{1}{2\pi}\int_{-\pi(1-\epsilon)}^{+\pi(1-\epsilon)}Xd(\lambda-\lambda_\Sun).
\label{average}
\end{equation}

Equations~\ref{a1}-\ref{o1} contain terms proportional to $\lambda^0$, 
$\sin(\lambda-\lambda_\Sun)$, $\sin(2\lambda-\varpi-\lambda_\Sun)$,
$\cos(2\lambda-\varpi-\lambda_\Sun)$, $\sin(\lambda-\Omega)$,
and $\cos(\lambda-\Omega)$. Inserting these factors into
Eqn~\ref{average} yields the following expressions:
\begin{equation}
<\lambda^0>=(1-\epsilon),
\end{equation}
\begin{equation}
<\sin(\lambda-\lambda_\Sun)>=0,
\end{equation}
\begin{equation}
<\sin(\lambda-\Omega)>=-\frac{\sin(\pi\epsilon)}{\pi}\sin(\Omega-\lambda_\Sun),
\end{equation}
\begin{equation}
<\cos(\lambda-\Omega)>=+\frac{\sin(\pi\epsilon)}{\pi}\cos(\Omega-\lambda_\Sun),
\end{equation}
\begin{equation}
<\sin(2\lambda-\varpi-\lambda_\Sun)>=+\frac{\sin(2\pi\epsilon)}{2\pi}\sin(\varpi-\lambda_\Sun),
\end{equation}
\begin{equation}
<\cos(2\lambda-\varpi-\lambda_\Sun)>=-\frac{\sin(2\pi\epsilon)}{2\pi}\cos(\varpi-\lambda_\Sun).
\end{equation}
The appropriate combination of these terms then yields the factors $d(\epsilon)$, $f(\epsilon)$
and $g(\epsilon)$ in Eqs~\ref{a1a}-\ref{o1a}.


\end{document}